\documentclass[12pt]{report}

\usepackage{amssymb}

\usepackage[T1]{fontenc}
\usepackage[
  variablett
]{lmodern}

\usepackage{listings}
\makeatletter
\lst@AddToHook{OnEmptyLine}{\addtocounter{lstnumber}{-1}}
\makeatother

\usepackage{verbatim} 
\usepackage{caption}
\usepackage{float}
\usepackage[top=2.5cm, bottom=2.5cm, left=3.5cm, right=3.5cm]{geometry}
\usepackage[frozencache,cachedir=_minted-main]{minted}
\usepackage[most]{tcolorbox}
\usepackage{tcolorbox}
\usepackage{listings}

\usepackage{tikz}
\usetikzlibrary{shapes.multipart}
\usetikzlibrary{arrows}
\usetikzlibrary{positioning} 
\usetikzlibrary{calc}

\newcommand{\HRule}{\rule{\linewidth}{0.5mm}}

\usepackage{listings}
\usepackage{xcolor}

\usepackage[unicode,pdfpagelabels,hidelinks]{hyperref}

\begin{document}
\pagenumbering{roman}
\setcounter{page}{1}


\lstset{basicstyle=\ttfamily,breaklines=true, numbers=left, 
numberstyle=\tiny, stepnumber=1, numbersep=3pt, numberfirstline=false, numberblanklines=false,
columns=spaceflexible}

\clearpage
\thispagestyle{empty}
\begin{center}
\vspace*{1cm}

{\huge\bfseries
Limited Read/Write-Set Hardware Transactional Memory\\
without modifying the ISA or the Coherence Protocol\par}
\vspace{1.5cm}

{\Large Konstantinos Kafousis\par}
\vspace{0.5cm}

\vfill

Computer Architecture \& VLSI Systems (CARV) Laboratory\\
Institute of Computer Science (ICS)\\
Foundation for Research and Technology -- Hellas (FORTH)\\[1cm]

{\Large\textbf{Technical Report}\par}
{\Large\textbf{FORTH-ICS/TR-496, August 2025}\par}
\vspace{2cm}

Work performed as a B.Sc Thesis at the\\
Department of Computer Science, University of Crete,\\
under the supervision of Manolis G.\,H. Katevenis, Panagiota Fatourou and Vassilis Papaefstathiou,
with the financial support of FORTH-ICS.\\[1cm]

\vfill
{\small Copyright 2025 by FORTH-ICS\par}
\end{center}
\clearpage

\clearpage
\thispagestyle{empty}
\begin{center}


\textsc{\LARGE Department of Computer Science}\\[0.5cm]
\textsc{\Large University of Crete}\\[1cm]

\textsc{\Large Diploma Thesis}\\[0.5cm]

\HRule \\[0.5cm]
{ \LARGE \bfseries Limited-Read/Write-Set}\\
\vspace{0.3cm}
{ \LARGE \bfseries Hardware Transactional Memory}\\ \vspace{0.3cm}
{ \LARGE \bfseries without modifying the ISA or}\\
\vspace{0.3cm}
{ \LARGE \bfseries the Coherence Protocol}\\
\vspace{0.2cm}

\HRule \\[1.3cm]

\begin{minipage}{0.4\textwidth}
  \begin{flushleft} \large
  \emph{Author:}\\
  Konstantinos \textsc{Kafousis}
  \end{flushleft}
\end{minipage}
\hspace*{\fill}
\begin{minipage}{0.4\textwidth}
  \begin{flushright} \large
  \emph{Advisors:}\\
  Manolis \textsc{Katevenis}
  \end{flushright}
\end{minipage}

\vspace{0.5cm}

\hspace*{\fill}
\begin{minipage}{0.4\textwidth}
  \begin{flushright} \large
  Panagiota \textsc{Fatourou}
  \end{flushright}
\end{minipage}

\vspace{0.35cm}

\hspace*{\fill}
\begin{minipage}{0.4\textwidth}
  \begin{flushright} \large
  \hspace{-0.17cm}Vassilis \textsc{Papaefstathiou}
  \end{flushright}
\end{minipage}

\vspace*{1cm}

\begin{center}
  \begin{minipage}{0.45\textwidth}
      \includegraphics[width=\textwidth]{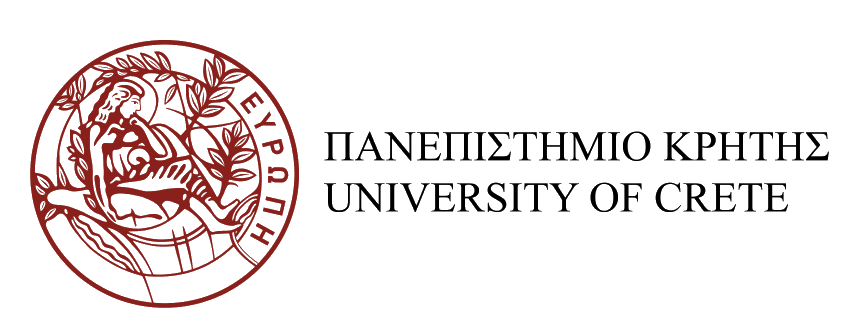}
  \end{minipage}
  \hspace{1cm}
  \begin{minipage}{0.45\textwidth}
      \includegraphics[width=\textwidth]{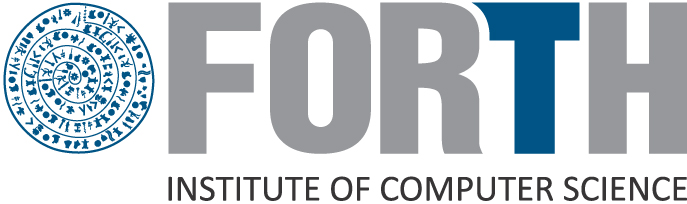}
  \end{minipage}
\end{center}

\vspace{2em}

\noindent\hspace*{-0.11\paperwidth}%
\begin{minipage}{0.9\paperwidth}
\centering
Work performed at the Computer Architecture and VLSI Systems (CARV) Laboratory \\
of the Institute of Computer Science (ICS), FORTH, Heraklion, Crete, Greece.
\end{minipage}

\vfill
{\large July 2025 \\ Updated Version - August 2025}

\end{center}
\clearpage

\clearpage
\thispagestyle{plain}    
\begin{center}
{\Huge \bfseries Abstract}
\end{center}
\vspace{1.5cm}

Hardware Transactional Memory (HTM) allows lock-free programming as easy as with
traditional coarse-grain locks or similar, while benefiting from the performance
advantages of fine-grained locking. Many HTM implementations have been proposed,
but they have not received widespread adoption because of their high hardware
complexity, their need for additions to the Instruction Set Architecture (ISA),
and often for modifications to the cache coherence protocol.

We show that HTM can be implemented without adding new instructions --- merely
by extending the semantics of two existing, Load-Linked and Store-Conditional.
Also, our proposed design does not modify or extend standard coherence
protocols. We further propose to drastically simplify the implementation of HTM
--- confined to modifications in the L1 Data Cache only --- by restricting it to
applications where the write set plus the read set of each transaction do not
exceed a small number of cache lines. We also propose two alternative mechanisms
to guarantee forward progress, both based on detecting retrial attempts.

We simulated our proposed design in Gem5, and we used it to implement several
popular concurrent data structures, showing that a maximum of eight (8) words
(cache lines) suffice for the write plus read sets. We provide a detailed
explanation of selected implementations, clarifying the intended usage of our
HTM from a programmer’s perspective. We evaluated our HTM under varying
contention levels to explore its scalability limits. The results indicate that
our HTM provides good performance in concurrent data structures when contention
is spread across multiple nodes: in such cases, the percentage of aborts
relative to successful commits is very low. In the atomic fetch-and-increment
benchmark for multiple shared counters, the results show that, under
low-congestion, our HTM improves performance relative to the Test-and-Test-and-
Set (TTS) lock.

\clearpage
\thispagestyle{plain}    
\begin{center}
{\Huge \bfseries Acknowledgments}
\end{center}
\vspace{1.5cm}

This thesis was conducted at the Computer Architecture and VLSI Systems
Laboratory (CARV) of the Institute of Computer Science (ICS) at the Foundation
for Research and Technology – Hellas (FORTH). I am deeply grateful to FORTH-ICS
for the financial support during the 2nd, 3rd, and 4th years of my studies
through the scholarships awarded to me.

At this point, I feel the need to express my sincere gratitude to my advisors,
for all they have done for me throughout this journey.

First of all, I am especially grateful to my advisor, Prof. Manolis Katevenis,
for his guidance, support, and inspiration. His advice has profoundly influenced
my thinking and writing.

I am also grateful to my advisor, Prof. Panagiota Fatourou, for her guidance,
support, and the constant, productive pressure she provided. \textit{"Write it
down,"} she used to tell me --- a phrase I will never forget.

Last but not least, I am grateful to my advisor, Prof. Vassilis Papaefstathiou,
for his valuable advice, to-the-point questions, and continuous support.

The guidance, support, and work ethic of all three of my advisors have been a
true example to me and have greatly contributed to both my academic and personal
growth.

Finally, I would like to thank my family for everything they have done for me.
They are the reason I am the person I am today.

\clearpage
\pagenumbering{arabic}
\setcounter{page}{1}

\newpage
\tableofcontents
\chapter{Introduction}

\section{Motivation}
Shared-memory multicore systems have become dominant in both consumer and high-performance computing platforms. These systems enable the parallel execution of multiple cooperating tasks that communicate through a common physical memory space. Parallel programming involves decomposing computation into multiple tasks that can execute concurrently, thereby reducing overall execution time. Designing such programs is challenging, as it requires careful synchronization and mechanisms for ensuring atomic access to shared data, in order to avoid race conditions and inconsistencies, while at the same time incurring minimal overhead to the performance of the parallel program.

As a result, there is an ongoing effort to develop techniques that improve both the programmability and performance of parallel programs. These techniques often introduce a trade-off between these two objectives. For example, traditional locking mechanisms (e.g., locks) have been widely used but offer limited performance, and reaching their performance potential typically demands significant design effort. On the other hand, mechanisms provided directly by hardware can offer better performance but come with certain drawbacks. For instance, hardware-supported atomic primitives are very limited in scope and can only support a narrow class of parallel programs. More flexible techniques, such as Hardware Transactional Memory (HTM)~\cite{HTMT} can support a much broader range of programs, but have required substantial architectural changes to be fully supported by systems. This thesis aspires to improve upon this latter point.

We argue that there is a need for a hardware mechanism similar to a general-purpose multi-word atomic primitive, or similar to a limited-scope Hardware Transactional Memory, without adding significant complexity to the system. To address this, we propose a hardware extension in the form of a limited-capacity HTM that covers the needs of a broader class of parallel programs than standard atomic primitives, while avoiding the design complexity associated with full-fledged general-purpose HTM implementations.

\section{Specific Problem}

Achieving atomicity in parallel programs is a non-trivial challenge, and numerous techniques have been developed in both software and hardware to address this issue. In this section, we present the main approaches, discussing the advantages and limitations of each.

A widely adopted solution is \textbf{locks}, although these come with notable overheads. On one hand, \textbf{coarse-grained locks} are easier for programmers to use correctly, but they hinder the full exploitation of parallelism, since a single lock protects an entire critical region of memory; as a result, a thread holding the lock blocks all other threads -- even those that access different, non-conflicting memory locations. On the other hand, \textbf{fine-grained locks} allow for greater concurrency, as multiple locks can independently protect different parts of memory; this enables threads operating on non-overlapping memory locations to proceed in parallel. However, fine-grained locking is considerably more difficult to program correctly. Programs using coarse-grained locks often end up executing almost sequentially, while those relying on fine-grained locks require greater design effort and programming expertise. Moreover, improper use may lead to concurrency issues such as deadlocks -where two or more threads wait indefinitely for each other to release locks- or starvation, where a thread is perpetually denied access to a lock and makes no progress, or priority inversion, where a lower-priority thread holds a lock needed by a higher-priority thread, delaying its execution.

Considering the limitations of traditional locking mechanisms, it becomes evident that there is a strong need for achieving atomicity without relying on locks. For this reason, most modern processor architectures provide hardware support for atomic primitives based on \textbf{read-modify-write (RMW) operations} on a single memory word. These have enabled the development of lock-free data structures and algorithms. Many of these have found practical application, relying solely on atomic operations onto a single word, often using compare-and-swap (CAS)~\cite{herlihysBook}. However, restriction to a single word introduces significant design effort and implementation complexity, which can ultimately limit performance. Furthermore, implementing complex concurrent data structures with only single-word CAS is difficult or even infeasible in many cases, due to its limited atomicity. Other approaches -particularly those that depend on atomic operations over multiple words- have not seen widespread adoption, largely due to the lack of hardware support.

To address the need for atomic read-modify-write operations involving multiple memory words, the concept of \textbf{Transactional Memory (TM)}~\cite{HTMT, STM} has been proposed as an alternative solution. This approach is inspired by the behavior of traditional database transactions, which allow multiple operations to be grouped and executed atomically. Transactional Memory (TM) allows programmers to define a block of code that must be executed atomically. By marking the boundaries of a transaction, they rely on the TM system to ensure that all memory operations within the transaction appear as equivalent to having been performed atomically -- that is, either all changes are applied or none. The TM system monitors all memory accesses performed during the transaction, tracks data dependencies across active transactions, and is responsible for updating memory with the new values if the transaction successfully commits, or restoring the old values if the transaction is aborted.

Approaches to implementing Transactional Memory have been proposed both in software --known as \textbf{Software Transactional Memory} (STM)~\cite{STM}-- and, in hardware -- known as Hardware Transactional Memory (HTM)~\cite{HTMT, logTM, TCC, UTM}. In the case of STM, several libraries have been developed to provide this functionality and simplify concurrent programming. However, STM is often less efficient than traditional synchronization techniques, such as locks, because transaction management and memory access tracking are handled entirely in software, introducing both computational and memory overhead that can significantly impact application performance.

In the case of \textbf{Hardware Transactional Memory (HTM)} --the area explored in this thesis-- the entire transactional mechanism is implemented directly in hardware, introducing minimal software and time overhead --but oftentimes considerable hardware complexity and cost. An HTM system must record the addresses read (read-set) and written (write-set) during a transaction, detect conflicts -which occur when two or more active transactions access the same memory location- and ensure that updates become visible to the rest of the system only if the transaction successfully commits. Otherwise, in the case of an abort, we must preserve the old values and restore the system to the exact state it was in prior to the transaction. These mechanisms are challenging both to reason about and to implement efficiently. As will be discussed in detail in the next section, most HTM proposals introduce significant architectural changes, including modifications to cache coherence protocols, extensions to the processor core, and additions to the instruction set architecture (ISA) in order to support fundamental transactional operations. This thesis aspires to improve upon this situation.

We propose a limited read/write-set Hardware Transactional Memory design that requires only minimal architectural modifications. Most importantly, we preserve the existing Instruction Set Architecture (ISA) by extending the semantics of load-linked and store-conditional instructions to serve as the transactional interface. We also utilize the standard cache coherence protocols (e.g., MESI) to detect conflicts and ensure atomicity, without requiring any modification or extension to them. Finally, we integrate a small set of Transactional Status Holding Registers (TSHRs) into the level-1 data cache to track the read/write-set.

\section{Contributions of this Thesis}

This thesis addresses the need for a low-cost Hardware Transactional Memory (HTM) that is highly compatible with existing hardware architectures.
\\\\
The advisors and the author of this thesis have contributed the following:
\begin{itemize}
    \item Proposed \textbf{a limited read/write-set Hardware Transactional Memory (HTM)} mechanism that:
    \item Requires \textbf{no new Instructions}, and only \textbf{extends} the semantics of two already existing instructions: load-linked and store-conditional.
    \item Requires \textbf{no modifications to standard cache coherence protocols}; and
    \item Introduces \textbf{low-cost hardware modifications} limited to the Level-1 data cache.
\end{itemize}

\noindent
The author of this thesis has contributed the following:
\begin{itemize}
    \item Proposed two alternative \textbf{hardware-only mechanisms that guarantee forward progress} under high-congestion scenarios.
\end{itemize}

\noindent
To analyze both the correctness and performance of the proposed hardware mechanism, the author of this thesis has contributed a complete simulation and evaluation of it. Specifically:

\begin{itemize}
    \item \textbf{Implemented the proposed design} on the \texttt{gem5} computer system simulator;
    \item \textbf{Developed custom microbenchmarks} to test and analyze the behavior of the mechanism under various conditions; and
    \item \textbf{Presents experimental results} evaluating performance and scalability.
\end{itemize}

\section{High-Level Comparison with Prior Work}

This section provides a high-level comparison between our proposed design and prior hardware-based approaches that aim to address the problem of atomicity and synchronization in parallel programs. A more extensive comparison is presented in Chapter 2.

\begin{itemize}
    \item \textbf{Read-Modify-Write (RMW) atomic primitives on a single memory address:} These primitives allow a thread to read a memory word, modify it, and write it back with the guarantee that no other thread will observe or interrupt the intermediate state~\cite{herlihysBook, intel_isamanual, arm_manual,riscv_manual}. Their key advantage lies in the fact that the operation is guaranteed to succeed in a single attempt. However, they are extremely limited in expressiveness, as they operate on only a single memory word. In contrast, \textbf{our design supports atomic execution over a group of multiple memory words}.

    \item \textbf{Double-Width Compare-and-Swap (DW-CAS) within a cache line:} This primitive allows a thread to atomically compare and swap two memory words, provided both reside within the same cache line~\cite{scott, intel_isamanual}. While more powerful than single-word primitives, this approach is still highly restrictive: it requires strict memory alignment and supports only one specific operation (compare and swap) over the values. \textbf{Our design does not require specific placement of memory words in memory, nor is it constrained to one, specific operation on them}.

    \item \textbf{Hardware Transactional Memory (HTM):} HTM allows threads to execute multiple memory operations as a single atomic transaction. However, existing HTM implementations~\cite{logTM, TCC, UTM, Bulk} require substantial architectural modifications, including ISA extensions (new, dedicated instruction(s)), modifications to cache coherence protocols, and significant hardware additions.
\end{itemize}

The first two categories are widely supported in modern processors and are therefore considered relatively trivial from a hardware implementation perspective. However, they only support a subset of parallel programs compared to our design. On the other hand, HTM mechanisms support a superset of cases relative to our design and have been the subject of extensive research, with many different schemes proposed in the literature. While some have been integrated into commercial systems, implementing a robust and efficient HTM mechanism remains a non-trivial task.
For this reason, in the next chapter, we provide a detailed overview of representative HTM implementations and highlight their core architectural differences compared to our design.

\section{Thesis Structure}

The remainder of this thesis is structured as follows:  
Chapter~2 discusses related work, focusing on proposed HTM designs.  
Chapter~3 presents the detailed architectural design of our proposed hardware mechanism.  
Chapter~4 describes the modifications made to the \texttt{gem5} simulator to describe and simulate our mechanism.  
Chapter~5 introduces the custom microbenchmarks developed for testing and analysis.  
Chapter~6 presents the evaluation results and performance analysis.  
Finally, Chapter~7 concludes the thesis with a summary of our contributions and directions for future work.

\chapter{Related Work}  

Several Hardware Transactional Memory (HTM) implementations~\cite{HTMT, logTM, TCC, UTM, Bulk} have been proposed in the literature, and some — such as those by Intel~\cite[pp.~1445--1457]{intel_cpp_guide} and IBM~\cite{ibm} — have even been integrated into commercial systems. As discussed in the previous chapter, HTM systems implement the entire transactional mechanism in hardware. In particular, they must track the set of memory locations read (read-set) and written (write-set) during the execution of a transaction, in order to support three critical functions: conflict detection, version management, and conflict resolution. This classification was first introduced in the design of LogTM~\cite{logTM} and was later thoroughly analyzed by Bobba et al.~\cite{bobba2007pathologies}, who demonstrated how different design choices across these functions can significantly impact the performance of HTM systems.

\textbf{Conflict detection} refers to the point in time at which the HTM system detects that a conflict has occurred. With \emph{eager conflict detection}, a conflict is detected as soon as a transaction:
\begin{itemize}
    \item writes to a memory location that is already in another transaction’s \textit{read-set} or \textit{write-set}, or
    \item reads from a memory location that is already in another transaction’s \textit{write-set}.
\end{itemize}
In contrast, \emph{lazy conflict detection} postpones conflict detection until a conflicting transaction attempts to commit.

\textbf{Version management} refers to how new and old values are stored during the execution of a transaction. These values arise from the writes performed during the transaction. In \emph{lazy version management}, old values remain in memory, while new values are temporarily stored in a write buffer; if the transaction commits successfully, the buffered writes are applied to memory. In contrast, \emph{eager version management} immediately writes the new values to memory, while storing the old values in a temporary buffer. This allows the system to restore the state it had prior to the transaction, in case the transaction is aborted.

\textbf{Conflict resolution} refers to the action taken when a conflict is detected between two or more transactions.  In the case of \emph{eager conflict detection}, the conflict must be resolved immediately, as soon as a transaction accesses a memory location that conflicts with another active transaction. The resolution policy may involve stalling the requester, aborting the requester, or aborting the other transaction(s).  With \emph{lazy conflict detection}, the conflict is typically detected at commit time, and the resolution policy may either abort all transactions that conflict with the committer, or choose to stall or abort the committer itself.

The behavior and performance of a HTM implementation depend on the particular combination of choices made across the three key operations. As also noted in the comprehensive analysis by Bobba et al.~\cite{bobba2007pathologies}, no single design point consistently outperforms the others across all workloads. Nevertheless, this classification provides a valuable framework for examining and comparing existing HTM proposals based on the design decisions they embody. In the remainder of this section, we review representative HTM systems, categorize them according to their conflict detection, version management, and conflict resolution strategies, and discuss the trade-offs associated with each combination. We conclude by highlighting the design decisions made in our own approach and the motivation behind them.

\textbf{Lazy conflict detection/ Lazy version management/ Committer wins}: In this category, each transaction stores its updates in a temporary write buffer. When it reaches the commit phase, it competes with other transactions that are also attempting to commit. The transaction that wins commit priority proceeds to commit and broadcasts its read-set and write-set. Any other transaction that detects a conflict with the committed transaction is aborted, while non-conflicting ones must attempt to gain commit priority again at a later time.
Two notable HTM systems that follow this approach are TCC~\cite{TCC} and Bulk~\cite{Bulk}. TCC proposes a complete replacement of traditional coherence and consistency protocols, and requires all programs to be written entirely in transactional form in order to execute on the TCC system - an aggressive but conceptually interesting design. Bulk, on the other hand, in addition to its HTM design, introduces a minimal mechanism to compactly encode the read-set and write-set of a committed transaction before broadcasting them - effectively reducing the overhead associated with this design decision during the commit phase.
This design configuration offers two significant advantages. First, it guarantees forward progress, as one transaction is always granted priority to commit in each attempt. Second, only the transaction that successfully commits has the authority to abort other conflicting transactions.
However, there are notable drawbacks as well. Only one transaction can commit at a time, even if multiple transactions are non-conflicting and have disjoint read/write sets. Additionally, due to the use of lazy conflict detection, many transactions may perform speculative work that ultimately gets discarded. The commit phase can also be long and costly, especially when the committing transaction has a large write-set (with write buffers typically ranging from 4 to 8 KB), which must be written back to memory and broadcast to the rest of the system.
Finally, these implementations require fundamental changes to the cache coherence protocols, making them difficult to integrate with existing hardware architectures.

\textbf{Eager conflict detection/ Lazy version management/ Requester wins}: In this category, each transaction stores its updates in a temporary write buffer and detects conflict eagerly - that is, as soon as a memory reference from another transaction accesses a memory location that is already part of the transaction’s read or write set. The transaction that detects the conflict must abort, while the requesting transaction continues execution and receives the memory response as normal. When a transaction completes, if no conflict has been detected during its execution, it writes its buffered updates to memory and commits.
A key advantage of this design is that it can leverage standard cache coherence protocols to perform conflict detection, which improves compatibility with existing hardware. However, it also introduces a significant drawback: a transaction that will eventually abort may still cause another transaction to abort prematurely. This can lead to pathological cases - particularly under high contention - where no transaction is able to make forward progress, a scenario clearly analyzed by Bobba et al.\cite{bobba2007pathologies} and also discussed in the analysis of Bulk\cite{Bulk}.
A notable implementation of this model is UTM\cite{UTM}, which supports transactions that may run for unbounded durations and have memory footprints exceeding the size of physical memory - a particularly interesting and ambitious feature. However, these capabilities add substantial complexity and require modifications to both the processor and the memory subsystem. In the same work, the authors also propose LTM, a more constrained model that reduces complexity, but still requires architectural changes to both the cache and the processor core.

\textbf{Eager conflict detection/ Eager version management/ Requester stalls}: In this category, each transaction writes its updates directly to memory and stores the old values in a temporary undo buffer. Conflicts are detected eagerly, as soon as a memory access from one transaction overlaps with the read or write set of another.
A representative implementation of this model is LogTM~\cite{logTM} and its variants~\cite{logTMse,nestedLogTM}, which stall the requester when a conflict is detected. This design introduces the risk of deadlock, as transactions may end up waiting on each other in cycles. To mitigate this, LogTM proposes a particularly interesting solution: it assigns a timestamp to each transaction and uses it to detect potential deadlocks — specifically, when a transaction that has stalled an older one would itself stall on another even older transaction. In such cases, the requester is aborted in order to break the cycle.
A simpler implementation following the same model is HTMT~\cite{HTMT}, in which the requesting transaction is immediately aborted upon conflicting with an active transaction. While this approach simplifies the design, it introduces the risk of forward progress violations under high-contention scenarios.
This design configuration offers a unique advantage due to the use of eager version management: the commit phase is very fast, as all writes have already been performed in memory. On the other hand, it introduces a corresponding drawback — aborting a transaction is expensive, since the system must restore all old values from the undo buffer. Finally, both HTMT and LogTM extend standard cache coherence protocols to support the required transactional functionality.

\begin{table}[ht]
\centering
\footnotesize
\begin{tabular}{|l|c|c|c|}
\hline
\textbf{System} & \textbf{Conflict Detection} & \textbf{Version Management} & \textbf{Conflict Resolution} \\
\hline\hline
\textbf{TCC}   & Lazy  & Lazy  & Committer Wins \\
\hline
\textbf{Bulk}  & Lazy  & Lazy  & Committer Wins \\
\hline
\textbf{UTM}   & Eager & Lazy  & Requester Wins \\
\hline
\textbf{LTM}   & Eager & Lazy  & Requester Wins \\
\hline
\textbf{LogTM} & Eager & Eager & Requester Stalls \\
\hline
\textbf{HTMT}  & Eager & Eager & Requester Aborts \\
\hline
\end{tabular}
\caption{Design choices of representative HTM systems}
\label{tab:htm-design-classification}
\end{table}

Beyond research proposals, HTM has also been adopted in certain commercial systems by Intel~\cite[pp.~1445--1457]{intel_cpp_guide} and IBM~\cite{ibm}.A thorough analysis of both implementations is provided by Nguyen in his master's thesis~\cite{nguyen_htm_thesis}, where he also investigates their transactional capacity limits - defined in terms of the number of loads and stores a transaction can contain. Nevertheless, neither implementation guarantees forward progress. For instance, Intel's HTM requires that a fallback code path be specified at the beginning of a transaction; if the transaction fails, control is transferred to that path, which typically involves acquiring a traditional lock.
Additionally, Intel’s x86 architecture supports a restricted two-word atomic primitive via a specialized instruction known as Double-Wide Compare-and-Swap (DW-CAS). This instruction requires that the two memory words be consecutive and reside within the same cache line - a constraint that significantly limits its practical usefulness.

While our work has been inspired by prior HTM proposals, it is important to clarify that our goal is not to design a complete HTM system, but rather to provide hardware support for a multi-word atomic primitive. Nonetheless, the problem we address closely resembles a restricted form of HTM, and therefore many of the trade-offs and architectural techniques used in full HTM systems are still relevant to our approach.

This constraint allows for several simplifications. For instance, challenges that arise when transactions exceed the size of the L1 data cache - a common source of complexity and cost in traditional HTM systems - do not apply in our case. Furthermore, rare events such as page faults, quantum expirations, or context switches may occur, but given their low probability, we chose not to implement dedicated hardware mechanisms for them. Instead, such events result in a transaction abort, helping us maintain minimal hardware complexity.

Having studied a wide range of HTM proposals, we adopt a design that aligns with the principles of simplicity and compatibility. Specifically, we choose eager conflict detection, and for conflict resolution, we rely on aborting the receiver, as this can be naturally supported by standard cache coherence protocols. We also adopt lazy version management, buffering updates until the transaction successfully commits. This decision is motivated by the observation that the transactions we target typically involve small write-sets, meaning that the commit phase - during which buffered writes are applied to memory - can be completed within just a few clock cycles.

In contrast to all prior HTM implementations discussed above, our design requires no extensions to the Instruction Set Architecture (ISA) to support transaction-related operations such as start, commit, or abort. Instead, it generalizes the semantics of the existing Load-Linked (LL) and Store-Conditional (SC) instructions in the RISC-V ISA to enable transactional behavior. Furthermore, the design requires no modifications to the processor core or the standard cache coherence protocol - not even changes to the content or structure of coherence messages. All transactional functionality is implemented through minimal additional hardware, restricted solely to the L1 data cache.

The complete architectural details of our proposed design are presented in the following chapter.

\chapter{Architectural Design}

In this chapter, we present in detail the proposed architectural design for our limited read/write-set Hardware Transactional Memory (HTM) system.

\section{Overview}

Our design builds upon typical \textit{shared-memory multiprocessor architectures}, where each processor has one or more \textit{private caches}, and coherence among them is maintained using a standard \textit{directory-based cache coherence protocol}.

The proposed implementation follows a \textbf{lazy version management} strategy (i.e., buffering transactional writes until commit) by introducing a small set of additional registers in the \textbf{L1 data cache}. These registers are used to:
\begin{itemize}
    \item Track the addresses of all cache lines accessed during the transaction (read-set and write-set).
    \item Store the new values written during the transaction into temporary line-size registers, which are made visible at the commit phase.
\end{itemize}

\noindent For conflict detection, we adopt an \textbf{eager conflict detection} (i.e., detecting conflicts as soon as they occur) approach that leverages the existing coherence protocols already present in most systems.

Unlike most conventional HTM implementations, we do \textit{not extend} the ISA with any new instruction. Instead, we \textit{extend} the \textit{semantics} of the already existing \texttt{load-linked} and \texttt{store-conditional} instructions, as follows:

\begin{itemize}
    \item The \textit{first} \texttt{load-linked} instruction issued by the processor (after the end of any previous transaction) \textbf{initiates a transaction}. 
    
    \item The \textit{single} \texttt{store-conditional} instruction \textbf{attempts to commit} the transaction, regardless of whether it accesses the same address as the first load-linked. Its success or failure indicates whether the transaction commits or aborts.
\end{itemize}

\noindent During the transaction (i.e., between the first \texttt{load-linked} and the single \texttt{store-conditional} instruction), the following semantics apply:

\begin{itemize}
    \item Subsequent \texttt{load-linked} instructions, as well as the first one, are treated as \textit{transactional loads}, and their accessed addresses are added to the transaction's \textit{read-set}.
    
    \item Regular \texttt{store} instructions executed within the transaction, as well as the single, terminating, store-conditional instruction are treated as \textit{transactional stores}; their target addresses and new values are added to the \textit{write-set}.
    
    \item Regular \texttt{load} instructions may also be executed within a transaction. These are considered \textit{non-transactional reads} and do \textbf{not} affect the transaction's read-set.

    \item The read-set and the write-set are kept track of at the granularity of entire cache lines -- not individual words, for reasons of simplicity of implementation. This way leads to unnecessary aborts in cases of false-sharing, but will not violate the atomicity properties of transactions.
\end{itemize}


\section{Hardware Extensions to the L1 Data Cache}

\begin{figure}[ht]
    \centering
    \includegraphics[width=1\textwidth]{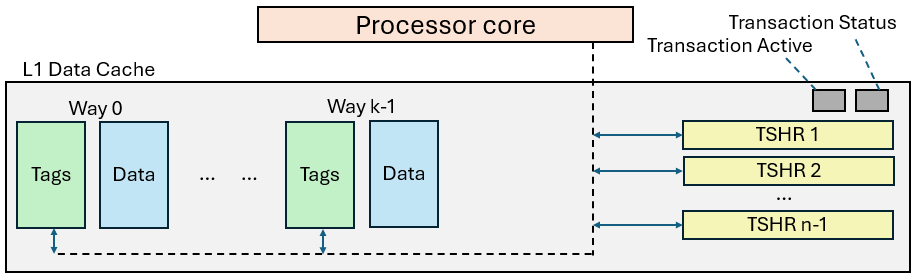}
     \caption{High-level view of the added state in the L1 Data Cache: right-hand side, in yellow and gray}
    \label{fig:dcache}
\end{figure}

Supporting transactional execution requires a mechanism capable of tracking the read-set and write-set of each transaction and compare them to those of other concurrent transactions executing on other processors in the same shared memory space. For this purpose, we introduce a small number of \textbf{Transaction Status Holding Registers (TSHRs)} in the L1 data cache as illustrated in Figure~\ref{fig:dcache}. The exact number and structure of these registers will be described in detail later in this section.

In most HTM implementations~\cite{logTM, TCC, UTM, logTMse, nestedLogTM}, similar functionality is achieved by adding extra metadata bits to each cache line in the L1 data cache. These bits typically indicate whether a given cache line belongs to the transaction’s read-set or write-set. Additionally, HTM systems commonly include a dedicated \textit{write buffer} to temporarily store either:
\begin{itemize}
  \item the new data to be written to memory upon successful commit (lazy version management), or
  \item the old data to be restored in case of an abort (eager version management).
\end{itemize}

\noindent In our design, which targets transactions with small read/write sets, we believe that adding a limited number of dedicated registers (TSHRs) is a more efficient and lightweight approach. Furthermore, we include a single \textbf{active Transaction bit} to indicate whether 
 there is an active transaction in progress.
 Most other HTM implementations typically integrate this bit directly into the processor core, as they are designed to handle more complex scenarios (e.g. context switch). In our case, however, the bit resides entirely within the L1 data cache, enabled by the simplifications made in our design. Moreover, an accompanying flag is used to indicate whether it has already been decided that currently active transaction will have to be aborted at its final commit phase.


\textbf{An inclusive policy is enforced between the L1 Data Cache and the TSHRs}. That is, any cache line tracked by a TSHR also resides in the L1 Data Cache. The only additional information stored in the TSHRs pertains to the \textit{write-set}, where the updated cache line data is buffered. In contrast, \textit{read-set} TSHRs do not store any data; instead, reads are performed directly from the contents of the cache.

\subsection{Transaction Status Holding Registers (TSHRs)}
Each Transaction Status Holding Register (TSHR) stores the following information:

\begin{itemize}
    \item \textbf{Tag of the cache line:} Identifies the cache line to which each memory word involved in a transactional read or write belongs. Conflict detection is therefore performed at \textit{cache line granularity}, like standard cache coherence protocols suggest.

    \item \textbf{Updated cache line data:} Stores the original cache line data as they were at the time of transaction start, modified according to the updated values for the write-set. These updates are applied to the main L1 data cache --and eventually to memory-- only if the transaction successfully commits.

    \item \textbf{Read/write-set bit:} Indicates whether the address associated with this TSHR belongs to the read-set or the write-set. If a line is accessed for both reading and writing, it is marked as part of the write-set.

    \item \textbf{Valid bit:} Specifies whether the TSHR is currently allocated to an active transaction and contains valid information.

   \item \textbf{Left-over bit:} Set when a transaction ends, for every TSHR that was valid during its execution. Although these entries are invalidated, the left-over bit marks them as recently used, allowing the next transaction to detect whether it accesses the same cache lines -- a hint that it may be a retrial of the previous (aborted) transaction.

\end{itemize}

\begin{figure}[ht]
    \centering
    \includegraphics[width=0.5\textwidth]{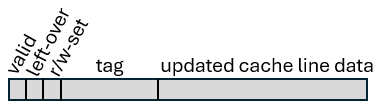}
    \caption{Fields of a TSHR}
    \label{fig:example}
\end{figure}

\textbf{The number of TSHRs added to the L1 Data Cache} determines the upper bound on the size of the read/write-set of transactions. As previously mentioned, each TSHR stores information for a single cache line, meaning that conflict detection operates at cache line granularity, as standard cache coherence protocols suggest.

Modern systems typically feature cache lines of 64 or 128 bytes, each capable of holding between 8 and 16 memory words assuming 64-bit word size. Based on this, the following observations can be made:

\begin{itemize}
    \item If two or more memory words from the read/write-set belong to the same cache line, a single TSHR will hold them. This scenario is common in parallel programs, as programmers often employ techniques such as data alignment and padding.
    \item In the general case, where each transactional word belongs to a different cache line, each such word will require a separate TSHR.
\end{itemize}

\noindent Taking the general case, we examine how many distinct cache lines can participate in the read/write-set of a single active transaction. This is ultimately constrained by the set-associativity of the L1 Data Cache (in modern processors, L1 data caches tend to be 8-way associative, which means that each cache set can hold up to 8 distinct cache lines mapped to the same index).

If the number of TSHRs exceeds the associativity, there is a risk that all addresses in the transaction map to the same cache index. In this case, some cache lines may be evicted prematurely, leading to transaction aborts that could have otherwise been avoided.

When the number of TSHRs is less than or equal to the associativity, this specific problem is avoided. However, the cache replacement policy may still evict a cache line belonging to the read/write-set in order to bring in a new one.

Commonly used policies such as Least Recently Used (LRU) and First-In First-Out (FIFO) are less likely to cause such issues, as cache lines belonging to the read/write-set are typically accessed frequently during the transaction and therefore remain recently used. In contrast, with a Random Replacement policy, such scenarios become more probable. In any case, a minor modification to the cache controller logic can fully prevent these situations.

\textbf{Based on the above analysis}, the following design considerations should be taken into account:

\begin{itemize}
    \item \textbf{The number of TSHRs should be equal to the associativity of the L1 Data Cache}(typically 8) to maximize transactional coverage and avoid index-based conflicts.
    
   \item The \textbf{cache controller} should be modified to \textbf{constrain the cache line replacement policy} in cases where \textbf{evicting a line} from the \textbf{read/write-set} could cause unnecessary transaction aborts.

\end{itemize}

\section{Conflict Detection}

In our design, we adopt the \textbf{eager conflict detection} approach at \textit{cache-line granularity}, leveraging the standard functionality of cache coherence protocols, without requiring any modifications to them.

A conflict between two or more transactions occurs when a cache line from the write-set of one transaction intersects with at least one cache line from the read-set or write-set of the other transaction, as illustrated in Table 3.1.

\renewcommand{\arraystretch}{1.2}
\begin{table}[h!]
\centering
\begin{tabular}{c||c|c}

 & Read\textsubscript{A}(X) & Write\textsubscript{A}(X) \\
\hline
\hline

Read\textsubscript{B}(X) & no conflict & conflict \\
\hline
Write\textsubscript{B}(X) & conflict & conflict \\
\hline
\end{tabular}
\caption{Conflicts on Cache Line X Between Transactions A and B}
\label{table:1}
\end{table}

\subsection{Cache Coherence Protocols}
\noindent In order to access a cache line, a processor must first acquire the appropriate coherence permissions:
\begin{itemize}
    \item To perform a \textbf{write} operation, the processor must have or obtain the cache line in the \textbf{Exclusive} state within its L1 data cache.
    \item To perform a \textbf{read} operation, the processor must have or obtain the cache line in either \textbf{Shared or Exclusive} state within its L1 data cache.
\end{itemize}

\noindent The coherence protocol is responsible for maintaining coherence when granting access to cache lines. Specifically:
\begin{itemize}
    \item Before granting \textbf{exclusive access} to a cache line to an L1 Data Cache, the protocol must first \textbf{invalidate all other copies} of that line in other caches or guarantee that no such other copies existed.
    \item Before granting \textbf{shared access} to a cache line, if another cache currently holds the line in an \textit{exclusive} state, the protocol must \textbf{downgrade that copy to shared} and ensure the most recent version of the data is forwarded to maintain consistency.
\end{itemize}

\begin{figure}[ht]
    \centering
    \includegraphics[width=1\textwidth]{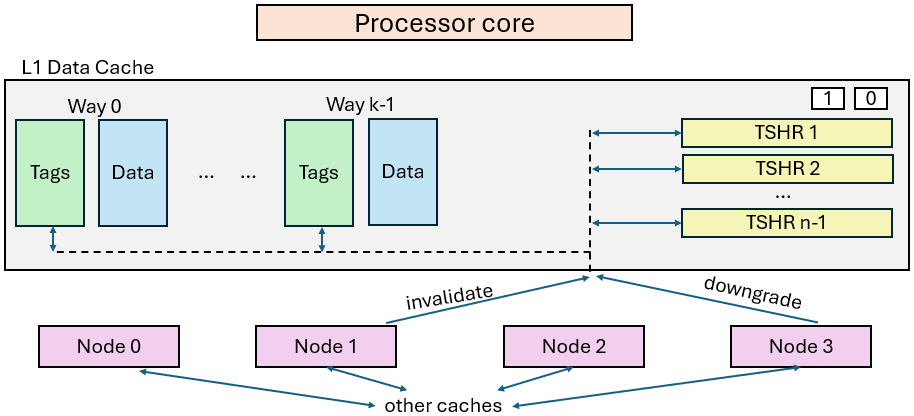}
    \caption{Detection of potential conflicts through incoming invalidate and downgrade requests from other cores/caches.}
\label{fig:conflict_detection}
\end{figure}

\noindent \textbf{Figure~\ref{fig:conflict_detection}} illustrates the requests arriving from coherence protocol that are used for detecting potential conflicts. Specifically, \textbf{when a transaction is active}, the addresses of the cache lines referenced by these requests are compared against the addresses tracked in the corresponding \textit{Transaction Status Holding Registers} (TSHRs). If any TSHR refers to the same cache line as the incoming request, a potential conflict may exist. More specifically, a conflict is detected in the following cases:

\begin{itemize}
    \item \textbf{Invalidate Request:} An invalidate indicates that another core intends to write to the cache line. In this case, a conflict exists \textbf{regardless of whether the cache line belongs to the transaction's \textit{read-set} or \textit{write-set}}.
    
    \item \textbf{Downgrade Request:} A downgrade indicates that another core wants to read the cache line, while the current cache holds it in the \textit{exclusive} state. A conflict exists \textbf{only if the cache line is part of the transaction's \textit{write-set} and has already been acquired in the \textit{exclusive} state}.
\end{itemize}

\section{Conflict Resolution}
For conflict resolution, we choose to \textbf{abort the receiver of the conflicting request}. According to the standard operation of cache coherence protocols, the receiver is the one that detects the conflict with another transaction based on the requests it receives from the directories (as explained in Section~3.3). 

\textbf{This design decision carries the risk} that transaction~A may abort transaction~B, while at the same time transaction~B may also abort transaction~A, leading to a \textbf{mutual abort}, even though one of the two transactions could have successfully committed.

This problem can be generalized to more than two transactions, where \textbf{each transaction aborts another} in a cyclic or cascading manner, preventing any of them from committing. As a result, the system may fail to make forward progress. In section 3.7, we propose a mechanism, upon repeated such transactional attempts, the hardware will recognize such a risk and modify its behavior so as to allow one of them to succeed.

\textbf{In all such cases, conflicts arise when a transaction either requests exclusive access to a cache line, or attempts to read a line that has already been granted exclusive access to another transaction.} This is the sole mechanism through which one transaction can cause another to abort. This observation is crucial in our implementation, where we pay particular attention to the point at which a transaction requests exclusivity for the cache lines it intends to write. This mechanism will be explained in detail in Section~\ref{sec:transaction_execution_flow}.


\section{Version Management}
In our design, we adopt the \textbf{lazy version management} approach. That is, writes performed during the execution of a transaction are not immediately applied to memory. Instead, they are temporarily stored in the \textit{TSHRs} as part of the transaction's write-set. These updates are only applied to memory when the transaction successfully commits. 

\begin{figure}[H]
    \centering
    \includegraphics[width=1\textwidth]{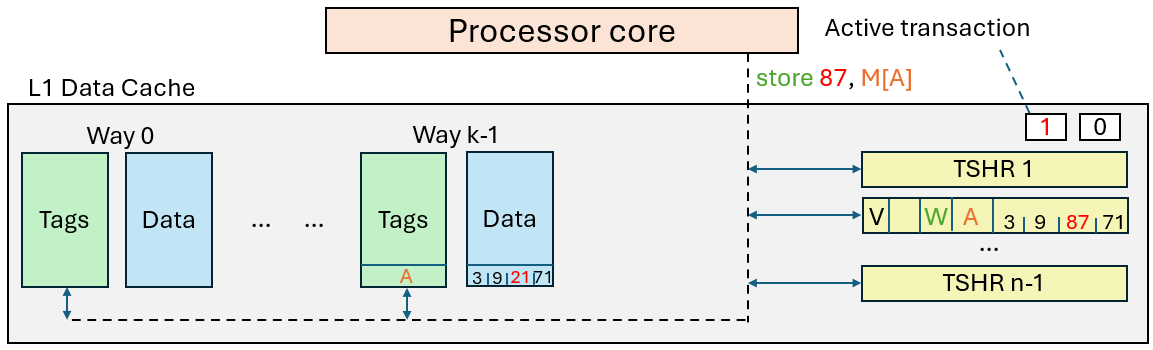}
    \caption{Recording a Store in the TSHR During Transaction Execution}
\label{fig:store}
\end{figure}

\textbf{Figure~\ref{fig:store}} illustrates the moment when a store reaches the L1 Data Cache while a transaction is active. In this case, the store operation allocates a free TSHR, sets its \textit{valid} bit, marks it as part of the \textit{write-set}, records the \textit{tag} of the address being written, and stores the new value.

\textit{Note:} The data field of the TSHR contains the entire cache line, but only the specific word targeted by the store is updated. No changes are made to the L1 Data Cache at this point. This TSHR entry remains active until the transaction either commits (in which case the updated data is written to L1 data cache) or aborts (in which case it is discarded).

\textbf{We chose this design decision based on the assumption that the write-set capacity is limited}. Therefore, applying the buffered writes to memory from the TSHRs at commit time --under the condition that the transaction already holds \textit{exclusive access} to all corresponding cache lines-- is a low-latency operation.

Specifically, if the number of TSHRs is 8 (which is likely the maximum due to the associativity of L1 data caches), then the read/write-set can hold at most 8 cache lines. In the worst-case scenario, where the write-set consists of 8 cache lines, committing the transaction requires writing back all 8 lines to memory. Since all cache lines are already in the L1 Data Cache in an exclusive state, this process can be completed in just 8 cycles —one per cache line— making the overhead negligible.

On the other hand, in the case of an \textit{aborted} transaction, no memory updates are performed, and thus the abort introduces no additional overhead on the memory side.

\section{Transaction Execution Flow Overview}
\label{sec:transaction_execution_flow}
This section provides a high-level overview of the transactional execution flow in the proposed mechanism. It outlines the key steps taken during the lifetime of a transaction, from its initiation to either commit or abort. The focus is on the sequence of operations performed by Load-Linked, Store, and Store-Conditional instructions, as well as their interaction with the TSHRs. In addition, a dedicated subsection discusses the design decision to defer the exclusivity request until the end of the transaction, along with the motivations and system-level benefits of this approach.

\noindent
The execution flow can be described through the following sequence of operations:

\renewcommand{\labelitemiii}{--}
\begin{itemize}
    \item A new transaction starts when a \textbf{Load-Linked (LL)} instruction is first executed.
    \item A transaction ends when a single \textbf{Store-Conditional (SC)} instruction is executed.

    \item During a transaction, each \textbf{Load-Linked (LL)} instruction:
    \begin{itemize}
        \item Compares its address to all TSHRs.
        \item If this is a new address, it is recorded in a new TSHR.
        \item Fetches or maintains its cache line in \textit{Shared} state.
        \item Adds the corresponding TSHR to the \textit{Read-Set}.
    \end{itemize}

    \item During a transaction, each \textbf{store} instruction:
    \begin{itemize}
        \item Compares its address to all TSHRs.
        \item If this is a new address, it is recorded in a new TSHR.
        \item Only tentatively modifies the data in the corresponding TSHR.
        \item Fetches or maintains its cache line in \textbf{Shared} state.
        \item Adds the corresponding TSHR to the \textit{Write-Set}.
    \end{itemize}

    \item The single and terminating \textbf{Store-Conditional (SC)} instruction:
    \begin{itemize}
        \item First performs all actions of a regular store as described above.
        \item If \textbf{no conflict} occurred during the transaction (i.e., no downgrade in the write-set and no invalidation in the read/write-set):
        \begin{itemize}
            \item The cache requests \textit{exclusive access} for \textbf{all cache lines in the write-set}.
            \item Once exclusivity is achieved and no conflict has been detected in the meantime, all tentative writes from the TSHRs are committed to the L1 Data Cache.
        \end{itemize}
        \item If a \textbf{conflict} arose during the transaction, the transaction is aborted.
    \end{itemize}
\end{itemize}

\begin{figure}[H]
    \centering
    \includegraphics[width=0.98\textwidth]{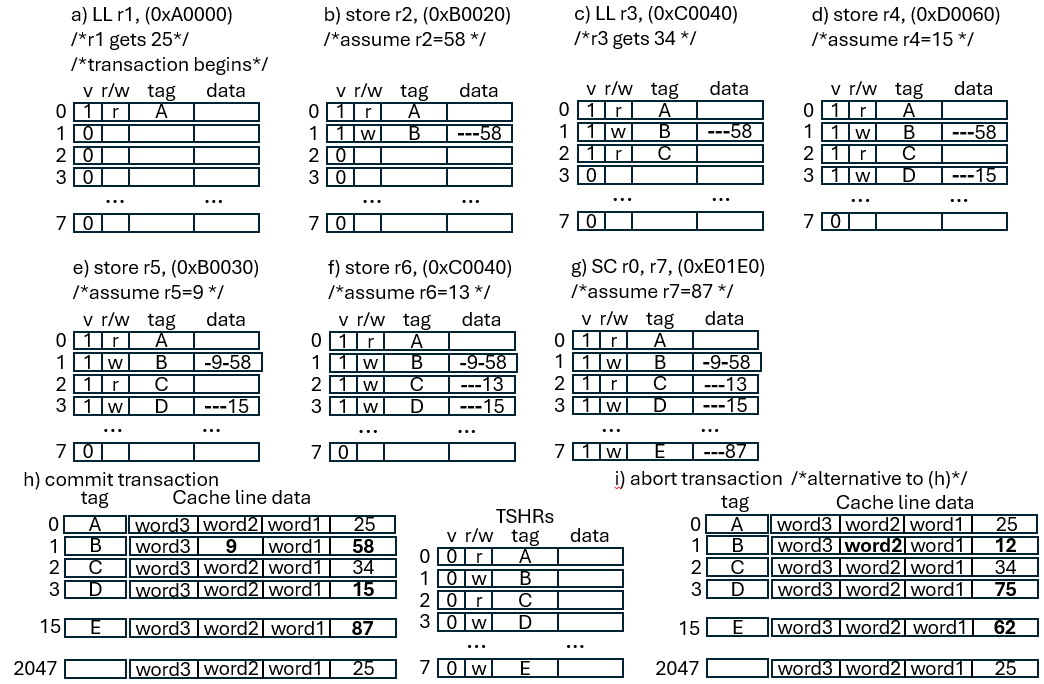}
    \caption{Transactional Execution Example with TSHR and Cache State Evolution
}
\label{fig:exec}
\end{figure}

\noindent
Figure~\ref{fig:exec} illustrates the state of the TSHRs and the L1 Data Cache during the execution of a transaction. For visualization purposes, we assume a 64-bit memory address space, a cache line size of 32 bytes ($2^5$), a word size of 8 bytes ($2^3$), and a direct-mapped L1 Data Cache of 65KB ($2^{16}$ bytes), resulting in a total of 2048 ($2^{11}$ ) cache lines.

\begin{figure}[H]
    \centering
    \includegraphics[width=0.5\textwidth]{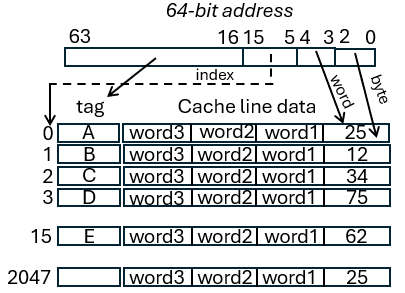}
    \caption{64-bit address breakdown and pre-transaction L1 data cache state
}
\label{fig:cache}
\end{figure}

\noindent
Figure~\ref{fig:cache} illustrates how a 64-bit memory address is decomposed into its constituent fields —tag, cache index, word offset, and byte offset— under the given configuration. It also shows the initial state of the L1 Data Cache before the transaction in Figure~\ref{fig:exec} begins.

For example, the address \texttt{0xB0020}, can be broken down into its constituent fields based on the system configuration defined above. In binary:

\[
\texttt{0xB0020} = 
\underbrace{\texttt{0000\ldots0000\ldots0000 1011}}_{\text{Tag (48 bits)}}
\underbrace{\texttt{000 0000 0001}}_{\text{Index (11 bits)}}
\underbrace{\texttt{00}}_{\text{Word}}
\underbrace{\texttt{000}}_{\text{Byte}}
\]

From this decomposition:
\begin{itemize}
    \item The lowest 3 bits (\texttt{000}) represent the byte offset
    \item The next 2 bits (\texttt{00}) represent the word offset
    \item The following 11 bits (\texttt{00..01}) represent the cache index, which equals 1
    \item The remaining most significant bits form the tag, which in this case is \texttt{0xB}.
\end{itemize}

\vspace{0.5em}

\noindent
\noindent
This figure~\ref{fig:exec} illustrates the contents of the TSHRs throughout the execution of a transaction, starting from its initiation in step~(a) and continuing until it either commits~(h) or aborts~(i). Figure~\ref{fig:cache} previously showed the initial state of the L1 Data Cache before the transaction begins.

\textbf{Note:} The contents of the L1 Data Cache remain unchanged throughout the transaction and are updated only at the end if the transaction commits. In addition, valid TSHRs that belong to the \textit{read-set} have an empty data field, since no speculative modifications are stored for those entries. For simplicity, the \textit{left-over bit} is omitted from this figure.

In the data fields of the TSHRs, each dash ``-'' represents one word within a cache line. For example, the notation ``- - - 13'' indicates that the value \texttt{13} is written into \textit{word 0} of the cache line, and that three more words follow it. This follows the little-endian byte ordering convention, in which bytes are arranged from least to most significant.

\noindent
\textbf{(a) The transaction begins:}  
The transaction begins as the first \textbf{Load-Linked (LL)} instruction reaches the L1 Data Cache. At this point, all TSHRs are assumed to be in the \textit{invalid} state. A TSHR is allocated and added to the \textit{read-set}. Additionally, the LL instruction activates the \textit{active transaction flag}, which is not shown in the figure for simplicity.

\noindent
\textbf{(b) A store accesses a new cache line (B):}
A \textbf{store} instruction reaches the L1 Data Cache, intending to write the value \texttt{58} into \textit{word 0} of cache line B. An \textit{invalid} TSHR (i.e., one that has not yet been used during the current transaction) is allocated and added to the \textit{write-set}. The store fills the TSHR with the data of the entire cache line, and only \textit{word 0} is updated with the new value. The remaining words remain unchanged, reflecting the values stored in the L1 Data Cache. Importantly, the L1 Data Cache itself remains unmodified at this point — even \textit{word 0} retains its original value — since updates are deferred until commit.

\noindent
\textbf{(c) A Load-Linked accesses a new cache line:}  
A second \textbf{Load-Linked (LL)} instruction accesses cache line C. An \textit{invalid} TSHR is allocated to track this address and is added to the transaction’s \textit{read-set}.

\noindent
\textbf{(d) A store accesses a new cache line:}  
A second \textbf{store} instruction reaches the L1 Data Cache, intending to write the value \texttt{15} into \textit{word 0} of cache line D. An \textit{invalid} TSHR is allocated and added to the transaction's \textit{write-set}. The TSHR is filled with the data of the entire cache line, and only \textit{word 0} is updated with the new value. The remaining words remain unchanged, reflecting the values stored in the L1 Data Cache.

\noindent
\textbf{(e) A store accesses a cache line already present in the write-set:}  
A third \textbf{store} instruction attempts to write the value \texttt{9} into \textit{word 2} of cache line B. Since a previous transactional store has already updated \textit{word 0} of the same cache line, a TSHR is already allocated and included in the \textit{write-set}. Therefore, no new TSHR is allocated—\textit{word 2} is simply updated in the existing TSHR. At this point, only \textit{words 3 and 1} in the TSHR still match the contents of cache line B in the L1 Data Cache.

\noindent
\textbf{(f) A store accesses a cache line already present in the read-set:}  
The fourth \textbf{store} instruction attempts to write the value \texttt{9} into \textit{word 0} of cache line C. A previous \textbf{Load-Linked (LL)} had already accessed this cache line, allocating a TSHR and adding it to the \textit{read-set}. As the cache line is already tracked by a TSHR in the \textit{read-set}, the corresponding bit is updated to indicate that this entry now belongs to the \textit{write-set}.
 At this point, the cache line data is fetched into the TSHR, and \textit{word 0} is updated with the new value. The remaining words retain the values stored in the L1 Data Cache.

\noindent
\textbf{(g) The single, terminating Store-Conditional arrives:}  
The single \textbf{Store-Conditional (SC)} instruction acts as a transactional store and then attempts to finalize the transaction, either by committing or aborting it. In this example, the SC accesses a new cache line E, allocates a new \textit{invalid} TSHR, and adds it to the \textit{write-set}. It fetches the full contents of cache line E from the L1 Data Cache into the TSHR, and updates only \textit{word 0} with the new value 87. The remaining three words retain the values stored in the L1 Data Cache.

If no conflict has been detected during the transaction, the SC then issues an \textit{exclusive access request} for all cache lines in the \textit{write-set} in parallel, in an attempt to commit the transaction.

\noindent
\textbf{(h) The transaction successfully commits:}  
At this point, no conflict had been detected when the \textbf{Store-Conditional (SC)} was issued, so an \textit{exclusive access request} was made for all cache lines in the \textit{write-set}. Furthermore, no conflict occurred in the interval between issuing the exclusivity request and actually acquiring exclusive ownership for all write-set lines. Therefore, the transaction is allowed to commit.

All tentative writes stored in the TSHRs belonging to the \textit{write-set} are now applied to the corresponding cache lines in the L1 Data Cache.

After the commit, all TSHRs are marked as \textit{invalid}, regardless of whether they were part of the read-set or the write-set. \textbf{Note:} the \textit{tag field} of each TSHR is not cleared—it remains intact, even though the entry is considered invalid. This design choice allows the next transaction to identify whether it is accessing the same cache lines as a previous transaction, thereby enabling the detection of repeated failures.

\noindent
\textbf{(i) The transaction is aborted:}  
In this case, which serves as the alternative to (h), a conflict was detected—either before the \textbf{Store-Conditional (SC)} instruction was reached (in which case no exclusivity was requested), or after the exclusivity request but before exclusive ownership was granted for all cache lines in the \textit{write-set}. As a result, the transaction is aborted.

No changes are made to the L1 Data Cache, and the cache lines that were tentatively updated during the transaction remain unchanged—their contents reflect the state prior to the transaction. 

As in the commit case, all TSHRs are marked as \textit{invalid}, regardless of their set membership, and their \textit{tag fields} are not cleared—they remain intact. This again allows the next transaction to identify whether it is accessing the same cache lines as the previous one, enabling detection of repeated failures.

\subsection{Exclusivity Request for the Write-Set}

Exclusive access for all cache lines in the \textit{write-set} is requested only at the end of the transaction, during the execution of the \textbf{Store-Conditional (SC)} instruction, and only if no conflict has been detected during the transaction.  
This design decision is guided by the principle of \textbf{lazy version management}, where updates are performed tentatively and only applied to memory if the transaction reaches commit.

Once exclusivity for the entire \textit{write-set} is acquired—and as long as no conflict is detected during this process—all tentative writes stored in the TSHRs are written to the L1 Data Cache. Each write operation is lightweight, typically completing in one cycle per cache line. In the worst-case scenario—when the write-set contains 8 lines—the entire commit phase completes in 8 cycles.

During the \textbf{commit phase} (i.e., once exclusive ownership has been acquired for all write-set lines), any incoming coherence requests (e.g., invalidates or downgrades) targeting cache lines in the read/write-set are stalled until commit writes complete.  
However, if a conflict is detected while attempting to acquire exclusivity—i.e., before reaching the commit phase—the transaction is aborted.

Delaying the exclusivity request until the execution of the single, terminating Store-Conditional instruction allows a transaction to avoid being prematurely aborted by other transactions whose commit point occurs earlier in time and that merely intend to read a cache line belonging to its write-set. 

\begin{figure}[H]
    \centering
    \includegraphics[width=0.8\textwidth]{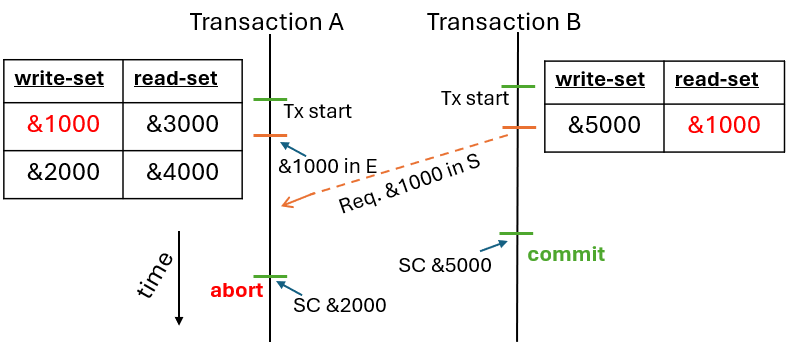}
    \caption{Acquiring exclusivity too early results in an abort due to a downgrade request from an earlier-committing transaction, while deferring the request would have permitted both transactions to commit.
}
\label{fig:req_excl_early1}
\end{figure}

This scenario is illustrated in Figure~\ref{fig:req_excl_early1}, where Transaction~A acquires exclusivity for a cache line in its write-set prematurely—immediately upon issuing a store. Transaction~B, which only intends to read that same cache line, triggers a downgrade request that causes Transaction~A to abort.

It is important to observe that, in this example, the commit point (i.e., the execution of the terminating SC instruction) of Transaction~B occurs before that of Transaction~A. Therefore, had Transaction~A deferred its exclusivity request until its own commit point, both transactions could have completed successfully without conflict.

However, if Transaction~B's commit point were to occur after that of Transaction~A, then a deferred exclusivity request from Transaction~A would rightfully abort Transaction~B. In that case, the abort is considered correct, since Transaction~A intends to write to the cache line, while Transaction~B only reads it.

\begin{figure}[H]
    \centering
    \includegraphics[width=0.8\textwidth]{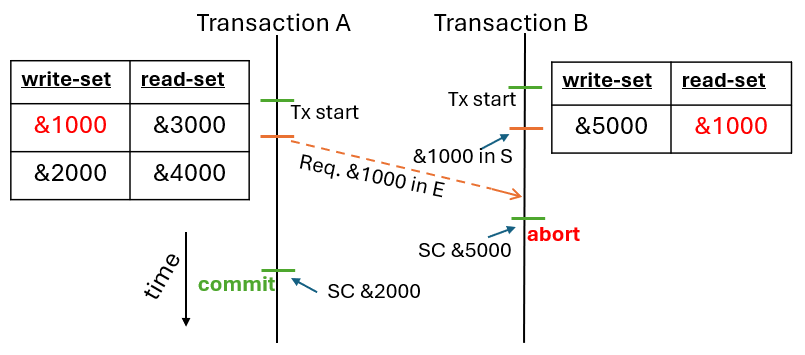}
    \caption{A premature exclusivity request causes the abort of another transaction, whereas deferring the request would have allowed both transactions to commit successfully.
}
\label{fig:req_excl_early}
\end{figure}

Conversely, issuing exclusivity requests too early may cause a transaction to abort other transactions that could have otherwise completed successfully. This effect is illustrated in Figure~\ref{fig:req_excl_early}, where Transaction~A acquires exclusivity prematurely—immediately upon issuing a store. As a result, it invalidates the corresponding cache line in Transaction~B, which merely intends to read it.

Again, we observe that since the commit point of Transaction~B occurs earlier than that of Transaction~A, both transactions could have committed without conflict if Transaction~A had deferred its exclusivity request until the execution of its store-conditional instruction.

Therefore, it is crucial to defer the exclusivity request for the write-set until the final phase of the transaction, specifically during the execution of the store-conditional instruction. This design choice offers several key advantages:

\begin{itemize}
    \item It increases the time window during which transactions whose read-set overlaps with the write-set of the current transaction can commit without receiving an invalidation.
    \item It reduces the time window during which a transaction is vulnerable to aborts caused by downgrade requests.
    \item It significantly limits unnecessary exclusivity requests, as such requests are issued at the end of the transaction and only if no conflicts have been detected up to that point. As a result, it increases the likelihood that exclusivity requests originate from transactions that are about to commit, rather than from those that will eventually abort.
\end{itemize}

\begin{figure}[H]
    \centering
    \includegraphics[width=0.7\textwidth]{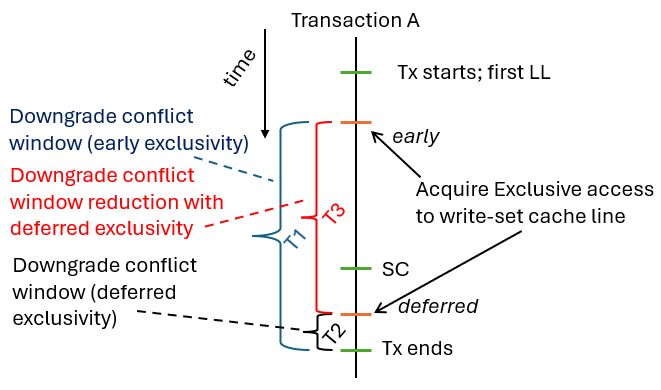}
    \caption{Impact of exclusivity acquisition timing on downgrade conflict exposure.
}
\label{fig:req_excl}
\end{figure}

These advantages are illustrated in Figure~\ref{fig:req_excl}, which compares the downgrade conflict windows resulting from early versus deferred exclusivity acquisitions.

As illustrated in the figure~\ref{fig:req_excl}, the interval $T_3$ represents both the reduction in the time window during which the current transaction is exposed to aborts due to downgrade requests, and the corresponding increase in the window during which other transactions—whose read-sets intersect with the current transaction's write-set—can successfully commit before being invalidated.

Additionally, the figure highlights why exclusivity requests are more likely to originate from transactions that will eventually commit. Specifically, the time window between the exclusivity acquisition and a potential conflict is significantly reduced—from $T_1$ to $T_2$—making it less likely that a transaction will acquire exclusivity and still be aborted shortly afterward. The difference between $T_1$ and $T_2$ corresponds precisely to $T_3$.

Beyond the advantages discussed above among active transactions, \textbf{minimizing the time window during which a transaction can be aborted due to downgrade requests is also important in scenarios involving non-transactional reads}.

A representative example is an optimistic search in a concurrent data structure, where traversal is performed without synchronization, and only upon reaching the target node is a validation step invoked. During the traversal phase, the search may read from a cache line that belongs to the write-set of an active transaction.

Thanks to the deferred exclusivity acquisition, if the active transaction has not yet reached its commit point (i.e., has not acquired exclusivity), the read does not trigger a conflict. In contrast, if the transaction has already acquired exclusivity for the cache line (i.e., it is at or beyond its commit point), the read rightfully causes the transaction to abort, preserving correctness.

Figure~\ref{fig:list} illustrates a concurrent doubly-linked list. At this point, the goal is not to describe how operations on such a concurrent data structure can be implemented using the proposed mechanism, but rather to emphasize the significance and frequency of non-transactional reads, and to highlight why it is important that such accesses do not cause unnecessary aborts in active transactions—especially when correctness is not violated.

In the example, we can imagine a transaction attempting to insert a new node with value 35 between nodes 30 and 40. As a result, its write-set includes the \texttt{next} pointer of node 30 and the \texttt{prev} pointer of node 40. Concurrently, another thread performs an optimistic search for node 60 and, during its traversal, reads the \texttt{next} pointer of node 30.

If the insertion transaction has not yet reached its commit point—meaning it has not acquired exclusivity for its write-set—the search traversal does not trigger a conflict and both operations can proceed concurrently. This demonstrates that deferring exclusivity acquisition allows the system to expose greater parallelism, without compromising correctness.

\begin{figure}[H]
    \centering
    \includegraphics[width=1\textwidth]{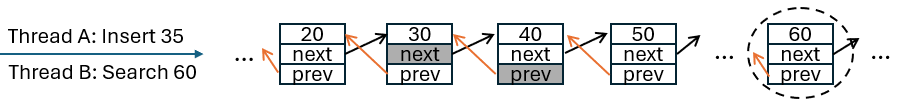}
    \caption{Concurrent access to a doubly-linked list with transactional insertion and non-transactional, optimistic search.
}
\label{fig:list}
\end{figure}

\section{Guaranteeing Forward Progress}

In Hardware Transactional Memory (HTM) systems that rely on \textit{eager conflict detection}, ensuring forward progress exclusively in hardware—regardless of the chosen conflict resolution policy—is a known, non-trivial challenge. This is because, when two transactions conflict, one of them must either \textit{stall} (risking deadlock) or \textit{abort} (risking livelock). 

As a result, many designs rely on a \textit{software contention manager} or adopt \textit{hybrid solutions}, where hardware attempts fast conflict resolution, but traps to software if the conflict persists. 

In this section, we describe two hardware-only techniques that guarantee forward progress without relying on software intervention: the \textit{Token-Based Priority} and the \textit{Sorted and Sequential Exclusivity Requests}.

In the \textit{Token-Based Priority} technique, a \textbf{single} transaction at a time is granted the token, and while it holds it, it stalls downgrade and invalidation requests targeting any cache line in its read/write-set. This prevents the transaction from being aborted due to conflicting accesses from other transactions and allows it to eventually commit.
 In contrast, the \textit{Sorted and Sequential Exclusivity Requests} technique allows \textbf{multiple} transactions to defer responding to such requests for a bounded time interval.

Both techniques are triggered upon detecting a repeated attempt and, importantly, are implemented entirely in hardware, requiring no changes or extensions to standard directory-based cache coherence protocol messages.

\subsection{Repeated Attempt}
This subsection defines the notion of a \textit{repeated attempt} and explains how it can be detected in hardware. Both techniques introduced later in this section for guaranteeing forward progress rely on identifying such repeated attempts.

A transaction may either commit successfully or abort due to conflicts or other events such as context switches. For this reason, programmers typically structure transactional code in a loop, so that in the event of an abort, the transaction is retried from the beginning until it eventually succeeds. An example of such a software pattern is shown in Figure~\ref{fig:while}, where the programmer aims to atomically increment the value at address \texttt{A} by~5 and the value at address \texttt{B} by~8, repeatedly executing the transaction until it succeeds and returns \texttt{1} in register \texttt{r0}.

\begin{figure}[ht]
\centering
\begin{ttfamily}
\begin{tabular}{l}
\textbf{do} \{ \\
\hspace{1em} load-linked \quad r3, M[A]; \quad // start transaction \\
\hspace{1em} load-linked \quad r4, M[B]; \\
\hspace{1em} add \quad r3, r3, 5; \\
\hspace{1em} add \quad r4, r4, 8; \\
\hspace{1em} store \quad r3, M[A]; \\
\hspace{1em} store-conditional \quad r0, r4, M[B]; \quad // commit transaction \\
\} \textbf{while} (!r0);
\end{tabular}
\end{ttfamily}
\caption{Software Pattern for Repeated Transactional Execution.}
\label{fig:while}
\end{figure}

As a result, a repeated attempt $k{+}1$ is likely to access the same memory locations as a previous attempt $k$, or at least a subset of the same read/write-set. To detect such cases, we introduce the \textit{left-over bit} in each TSHR. When a transaction completes (either by committing or aborting), we do not clear the \textit{tag} field of its TSHRs. Instead, we set the \textit{left-over bit} in each TSHR that was valid at the end of the transaction.

Then, for every new Load-Linked or Store instruction in a subsequent transaction, the tag of the accessed address is compared against the tags of all TSHRs with the left-over bit set. If a match occurs, a counter is incremented. When this counter reaches a predefined threshold—or if the number of matches covers a certain percentage of the currently active read/write-set—we infer that the current transaction is likely a repeated attempt.

\begin{figure}[H]
    \centering
    \includegraphics[width=0.7\textwidth]{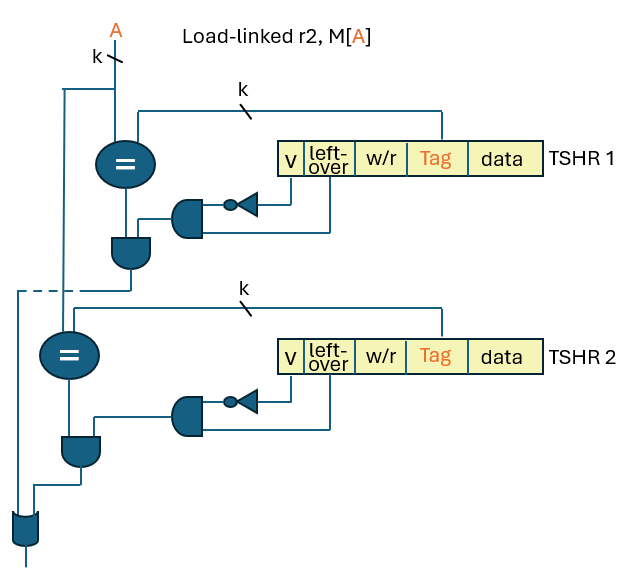}
    \caption{Tag comparison logic for identifying repeated transactional attempts using left-over TSHRs.}

\label{fig:leftover}
\end{figure}

Figure~\ref{fig:leftover} illustrates a simple hardware circuit for tag comparison involving two TSHRs. A Load-Linked instruction arrives, accessing an address with tag~\texttt{A}, which is then compared in parallel against the tags of all TSHRs that are both invalid and have their \textit{left-over} bit set. The circuit produces a one-bit output that signals whether the incoming tag matches any of the stored tags in left-over TSHRs.

In such cases, when a potential repeated attempt is detected, the system can trigger a dedicated action to assist this transaction—or at least one of the conflicting ones—in making progress, thereby ensuring that the system avoids entering a state of livelock or deadlock.

\subsection{Token-Based Priority}
This technique is inspired by the classic idea of a \textit{token}, where the entity holding the token is granted priority. In our implementation, the token is represented by a specific memory address—an entire cache line—and a transaction obtains the token by bringing this cache line into its L1 Data Cache in the \textit{Exclusive} state.

More specifically, when a transaction is identified as a repeated attempt and no conflict or abort-triggering event has yet occurred, it issues a request to acquire the token cache line in \textit{Exclusive} state and \textbf{continues executing without waiting for the token to be granted}. If it successfully receives the token and no abort condition has been triggered in the meantime, the transaction proceeds and enters a privileged mode in which it delays all invalidation requests for any cache line in its read/write-set, as well as downgrade requests for its write-set and downgrades/invalidations for the token itself. These requests are stalled until the transaction completes and commits.

This approach has three key advantages:  
\begin{itemize}
    \item Only a single L1 Data Cache delays coherence responses at any given time, ensuring that no deadlock can arise.
    \item All non-conflicting transactions are still allowed to complete concurrently with the token-holder, including those that requested the token but managed to commit before acquiring it.
    \item Any coherence stalls are caused only by the transaction that currently holds the token, which is expected to eventually commit unless interrupted by an external event such as a context switch.
\end{itemize}

However, the main drawback is that the token is centralized. Therefore, in scenarios with many simultaneous repeated attempts, contention may build up at the directory node responsible for the token cache line (i.e., its home node), potentially creating a bottleneck.

\begin{figure}[H]
    \centering
    \includegraphics[width=0.6\textwidth]{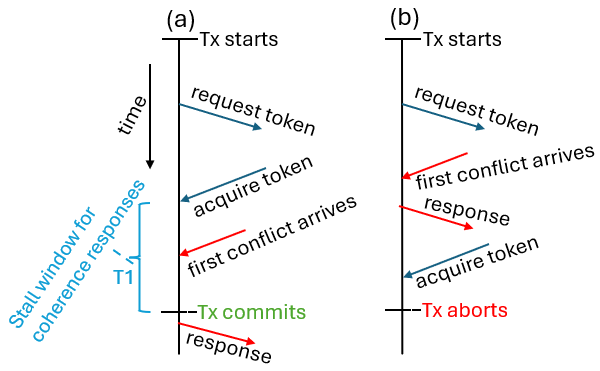}
    \caption{Impact of token acquisition timing on coherence request handling.}

\label{fig:token}
\end{figure}
\noindent
Figure~\ref{fig:token} illustrates two scenarios involving the timing of token acquisition relative to the arrival of coherence requests targeting the transaction's read/write-set.

\noindent
In case (a), the token is acquired \textit{before} the first invalidation or downgrade request for the read/write-set arrives. As a result, the L1 Data Cache delays responding to any coherence request it receives during interval~$T_1$ and the transaction eventually commits.

\noindent
In case (b), the token is acquired \textit{after} an invalidation or downgrade request for the read/write-set has already arrived and been served by the L1 Data Cache. Consequently, the cache continues to respond normally to all coherence requests, and no stalling occurs, since the transaction will eventually abort.

\noindent
Note that in case (b), the token is acquired just before the transaction aborts and returns failure. In some cases, the token might even be received shortly after the transaction has already ended. Since the software will likely reattempt execution until the transaction successfully commits, it may be beneficial to check at the beginning of a transaction whether it already holds the token—effectively granting it immediate priority—rather than waiting to be classified as a repeated attempt.

\begin{figure}[H]
    \centering
    \includegraphics[width=0.6\textwidth]{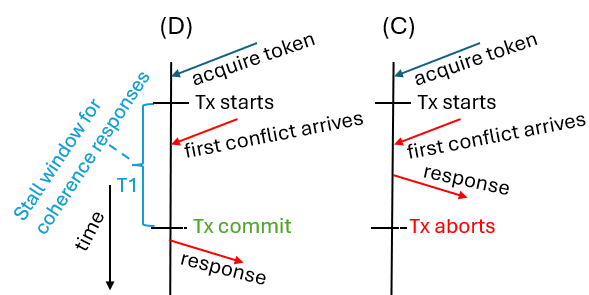}
    \caption{Effect of Early Token Check on Transaction Success.}

\label{fig:token2}
\end{figure}
\noindent
Figure~\ref{fig:token2} illustrates two scenarios in which a transaction holds the token from the beginning—not because it explicitly requested it, but because a previous attempt had requested the token, acquired it too late, and eventually aborted. Since no other transaction requested the token in the meantime, the next attempt begins execution while already holding it.

\noindent
In scenario (c), the transaction performs an early check at the beginning of its execution and detects that it already holds the token. As a result, it stalls all coherence requests throughout its lifetime, maintains priority, and successfully commits.

\noindent
In scenario (d), no such early check is performed. Although the transaction holds the token from the start, it fails to take advantage of it and eventually aborts due to a coherence conflict that could have been avoided.

Considering that the time interval between the termination of a transaction and the start of its repeated attempt is typically very short\footnote{This delay typically involves a simple comparison on a return flag to determine whether the transaction succeeded, and in some cases may include a short back-off time inserted by the programmer to reduce contention.}—just a few clock cycles—it is quite likely that the token remains in the L1 Data Cache. Therefore, it may be beneficial to perform a token presence check at the beginning of each transaction. This check amounts to a simple tag lookup in the L1 Data Cache, requiring only a single clock cycle, and can immediately grant the transaction execution priority if the token is found.

\begin{figure}[H]
    \centering
    \includegraphics[width=0.4\textwidth]{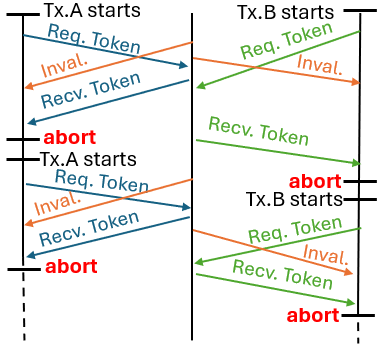}
    \caption{A pathological scenario where all transactions acquire the token after conflict detection, resulting in no progress.}

\label{fig:infinity_aborts}
\end{figure}

\noindent\textbf{Conclusions on Token-Based Priority.}  
It is crucial that a transaction is identified as a repeated attempt early in its execution, so that it has enough time to request and acquire the token before detecting its first conflict. Otherwise, a pathological case may occur—as illustrated in Figure~\ref{fig:infinity_aborts}—where all transactions acquire the token only after having already observed a conflict, and thus none of them ever make forward progress.

To address this issue, we propose two possible enhancements:

\begin{enumerate}
   \item \textbf{Sequential Abort Counter Heuristic:} Each L1 Data Cache maintains an $n$-bit counter that is incremented every time a transaction aborts and reset upon a successful commit. When a new transaction begins\footnote{The $k$-th attempt of Transaction~$T_1$ may be executing on processor core $P_1$, while the $(k{+}1)$-th attempt of the same transaction may run on a different core $P_2$. This is not problematic, as under high-contention scenarios the goal is to guarantee forward progress for at least one transaction—not necessarily the one with the most prior attempts.}, (i.e., with the first \texttt{Load-Linked} instruction), it checks the value of this counter.
 If the counter has reached its maximum value ($2^n{-}1$), the transaction proactively issues a token request \textit{before} accessing any cache line in its read/write-set. It then waits for a fixed number of clock cycles to acquire the token. If the token is not received within this time window, the transaction proceeds with normal execution to avoid potential deadlock.

    \item \textbf{Software-Controlled Token Acquisition:} The programmer may explicitly initiate early token acquisition by performing the first \texttt{Load-Linked} on a special reserved address, prior to starting the transactional region. This resembles a soft form of lock acquisition, though it remains non-blocking, as all non-conflicting transactions may still commit concurrently.
\end{enumerate}

\noindent

\subsection{Sorted and Sequential Exclusivity Requests}

In the baseline mechanism described so far, the \texttt{Store-Conditional} instruction—
provided no conflict has been detected—triggers parallel exclusivity requests for all cache lines in the transaction’s write-set.

In contrast, the technique we now describe modifies this behavior specifically for repeated attempts. When a repeated attempt reaches its \texttt{Store-Conditional} instruction and no conflict has been detected, the write-set is first sorted in increasing order based on (virtual or physical)\footnote{The choice depends on whether the mechanism is intended to support only multi-threaded programs within the same process (in which case virtual addresses suffice), or multiple processes communicating through shared memory, which requires physical address ordering.} address.
Then, exclusivity is requested sequentially—one cache line at a time. That is, the transaction issues a request for the first cache line, and only after it has been granted exclusive access, it proceeds to request the next one, and so on.

During this phase, if an \texttt{Invalidate} arrives targeting either any cache line in the read-set, or a cache line in the write-set that has not yet been acquired in \textit{Exclusive} state, the transaction is immediately aborted.
 However, if an \texttt{Invalidate} or \texttt{Downgrade} request targets a cache line in the write-set that has already been acquired in \textit{Exclusive} state, the L1 Data Cache delays responding for a predefined number of clock cycles. This temporary stalling provides the transaction with an opportunity to complete its sequential exclusivity acquisitions and commit successfully. Additionally, it prevents multiple transactions from indefinitely delaying coherence responses. As a result, the system avoids circular waiting and breaks potential deadlock scenarios.

\begin{figure}[H]
    \centering
    \includegraphics[width=1\textwidth]{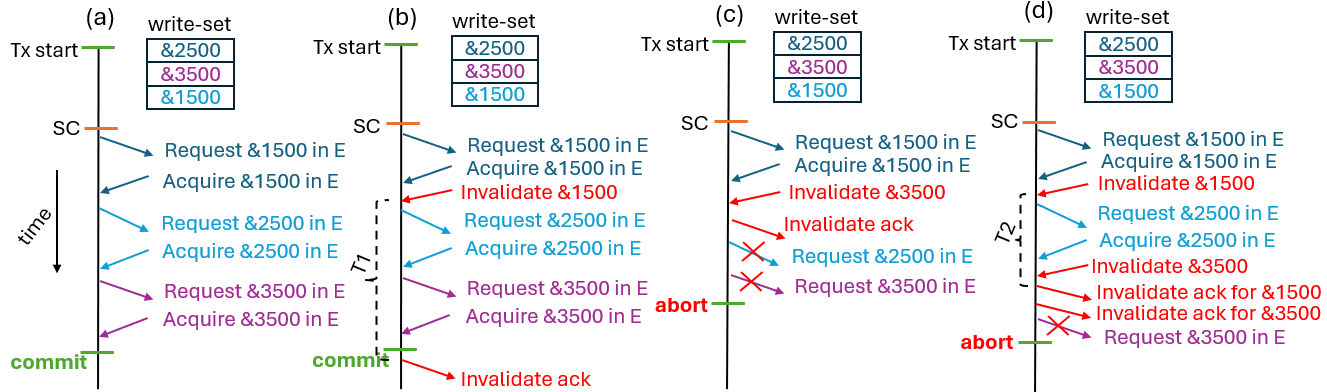}
    \caption{Sequential Exclusivity and Coherence Request Handling in Repeated Attempts
}

\label{fig:request_in_order}
\end{figure}

Figure~\ref{fig:request_in_order} illustrates several representative scenarios that may occur when using the sequential exclusivity mechanism for repeated attempts.

\noindent
In (a), the transaction sequentially requests and successfully acquires exclusive access to all cache lines in its write-set, completing without encountering any coherence requests and committing successfully. 

\noindent
In (b), the transaction acquires the first cache line (1500) and subsequently receives an \texttt{Invalidate} for it. The response is delayed for a time interval denoted as $T_1$, which corresponds to the remaining duration of the transaction's execution. During this time, the transaction acquires exclusivity for the rest of the write-set and eventually commits successfully.

\noindent
In (c), an \texttt{Invalidate} targeting the third cache line (3500) in the write-set arrives before that line is acquired. Since the line has not yet been acquired in \textit{Exclusive} state, the transaction is aborted immediately without issuing further exclusivity requests.

\noindent
In (d), an \texttt{Invalidate} arrives for the first cache line (1500) after it has already been acquired in \textit{Exclusive} state. The L1 Data Cache delays its response to this request. However, a subsequent \texttt{Invalidate} targeting the third cache line (3500) arrives before that line has been acquired. As a result, the transaction is immediately aborted and all previously stalled coherence requests are responded to at that point.

\textbf{By issuing exclusivity requests for the write-set in a sorted and sequential manner, there will always be at least one transaction that does not encounter conflicts in its write-set.} This eliminates circular contention, ensuring that it is not possible for Transaction~T$_1$ to abort Transaction~T$_2$, Transaction~T$_2$ to abort Transaction~T$_3$, and so on, up to Transaction~T$_{n}$ aborting Transaction~T$_1$.

\begin{figure}[H]
    \centering
    \includegraphics[width=0.9\textwidth]{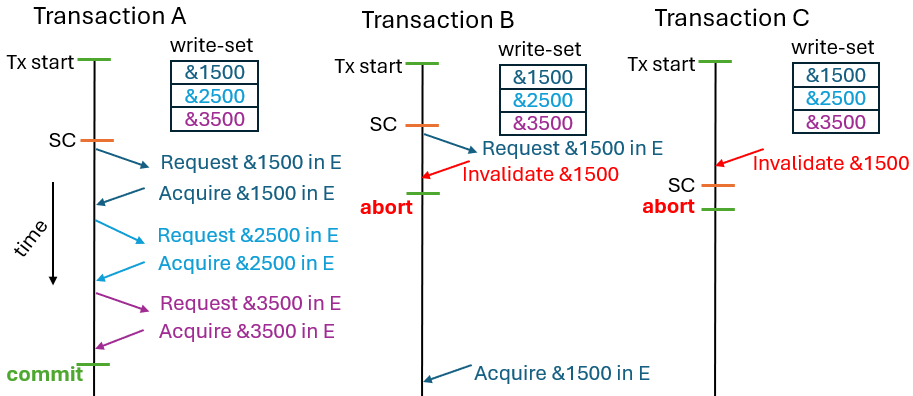}
   \caption{Transactions~A and B both issue requests for exclusive access to the first cache line (1500), but the request from Transaction~A reaches the directory first and is granted. As a result, Transactions~B and C receive \texttt{Invalidate} messages for cache line 1500 and abort. In the case of Transaction~C, the \texttt{Invalidate} arrives before the \texttt{Store-Conditional} instruction, preventing it from issuing any exclusivity requests at all. Transaction~A proceeds to acquire the rest of its write-set and commits successfully. 
  \textit{Note:} If Transactions~B and C are re-executed and reissue either exclusive or shared coherence requests for address 1500 while Transaction~A has not yet committed, Transaction~A will not be aborted, as its L1 Data Cache will postpone responding to such requests until the commit completes.
}
\label{fig:same_write_set}
\end{figure}

Figure~\ref{fig:same_write_set} illustrates such a case with three active transactions, all sharing the same write-set. The transaction that first acquires the initial cache line in \textit{Exclusive} state will proceed to complete successfully—in this case, Transaction~A.

\textit{Note:} Other transactions, such as Transaction~B, may also issue requests for the same cache line before receiving an \texttt{Invalidate}, but only one request will ultimately reach the directory first via the interconnection network and receive priority.

Even in more general scenarios, where the write-sets of active transactions are not exactly identical but may partially overlap, there will still always be at least one transaction that does not encounter a conflict in its write-set. 

To prove this, consider an attempt to construct a circular dependency involving three active repeated transactions, where each transaction aborts the next due to a conflict on a different cache line. Let Transaction~T$_1$ abort Transaction~T$_2$ due to address $X$, Transaction~T$_2$ abort Transaction~T$_3$ due to address $Y$, and Transaction~T$_3$ abort Transaction~T$_1$ due to address $Z$.

\noindent
Because exclusivity requests are issued sequentially and in increasing address order, this implies that:

- Transaction~T$_2$ requested $Y$ before acquiring $X$, hence $Y < X$

- Transaction~T$_3$ requested $Z$ before acquiring $Y$, hence $Z < Y$

- Transaction~T$_1$ requested $X$ before acquiring $Z$, hence $X < Z$

\noindent
However, combining the above inequalities gives $Y < X < Z < Y$, which is a contradiction. Therefore, it is not possible for all three transactions to mutually abort each other through conflicts on their write-sets. 

This logic extends to any number of transactions, proving that the sequential and ordered exclusivity acquisition policy inherently avoids deadlocks and guarantees that at least one transaction will not be aborted due to a conflict in its write-set.

In the analysis above, we proved that by issuing exclusivity requests for the write-set in a sorted and sequential manner, at least one of the repeated attempts will not encounter a conflict in its write-set. This conclusion assumes that all conflicting active transactions are, in fact, repeated attempts that follow the sequential exclusivity acquisition policy.

If, however, even a single transaction is not a repeated attempt and instead issues parallel exclusivity requests for its entire write-set, the above property may no longer hold. Nevertheless, this does not pose a correctness issue. In high-contention scenarios, if no transaction manages to complete, all active transactions will eventually become repeated attempts through retry mechanisms. If not, then at least one transaction must have committed, allowing a new transaction to begin—thus forward progress is preserved in either case.

It is important to note, however, that the analysis above guarantees that at least one transaction will not be aborted due to conflicts in its write-set. It does not address potential conflicts involving the read-set. Even if all active transactions have disjoint write-sets, they may still cause each other to abort due to interactions between the read-set of one and the write-set of another.

\begin{figure}[H]
    \centering
    \includegraphics[width=0.6\textwidth]{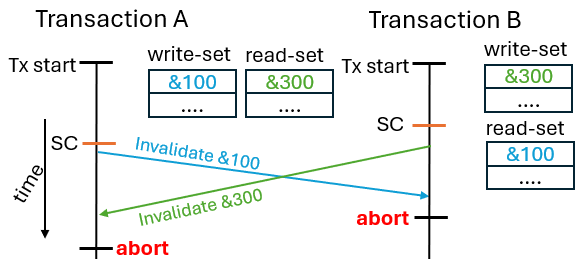}
    \caption{illustrates such a scenario, where the write-sets of two transactions are disjoint, but there exists an overlap between the read-set of one and the write-set of the other. As a result, both transactions are eventually aborted, and neither is able to commit—despite the fact that exclusivity requests are issued in a sorted and sequential manner. This example highlights that write-set ordering alone is not sufficient to prevent conflicts when read/write-set interactions are present.
}

\label{fig:mutual_abort}
\end{figure}

\noindent\textbf{Conclusions on Sorted and Sequential Exclusivity Requests.}
This technique guarantees forward progress in scenarios where only the write-sets of active transactions conflict, while their read-sets remain entirely disjoint. Therefore, it is not sufficient to guarantee forward progress in general-purpose transactional workloads. 

Nevertheless, it effectively covers a significant class of applications where transactions are used to support \texttt{multi-word read-modify-write} operations, such as multi-word \texttt{compare-and-swap} or \texttt{fetch-and-add}. In such use cases, the read-set eventually becomes the write-set, and by the time exclusivity is requested, the read-set is effectively empty—making this technique an ideal fit.

Moreover, since in our design the transactional write-set size is limited, issuing exclusivity requests sequentially rather than in parallel does not impose significant additional overhead. Importantly, this overhead is only paid by transactions that are repeated attempts—i.e., in situations of contention where forward progress must be ensured. All other non-conflicting transactions complete quickly without incurring this cost.

\section{Transactional Execution Lifecycle}

In the previous sections, we presented all the techniques and design decisions that constitute our implementation. This section serves as a summary and integration point for those concepts, detailing the distinct phases in the lifetime of a transaction.

A transaction—leveraging the token-based priority technique—progresses through a series of well-defined states:

\begin{itemize}
    \item \textbf{Outside Transaction:} No transaction is currently active. Load and store instructions execute normally, without any transactional tracking.

    \item \textbf{Inside Transaction – Status: Non-Abort:} A transaction is currently active, and no abort-triggering event has occurred. \textit{Note:} The L1 Data Cache may have received coherence requests for cache lines in the transaction's read/write-set, but it defers responses if the transaction holds the token.

    \item \textbf{Inside Transaction – Status: Abort:} A transaction is active, but a conflict or other abort-triggering event has occurred. The transaction is marked for abort, and the L1 Data Cache immediately responds to all coherence requests without deferral.

    \item \textbf{Exclusivity Acquisition Phase:} The terminating \texttt{Store-Conditional} instruction is issued, and the transaction has not been marked for abort. The cache issues requests for \textit{exclusive access} to all cache lines in the write-set.

    \item \textbf{Commit Phase:} All exclusivity requests are granted, and no abort-triggering event has occurred. During this phase, the L1 Data Cache defer all coherence responses. Once the speculative updates from the TSHRs are written back to the L1 Data Cache, all TSHRs are invalidated, a commit signal is sent to the processor, and the transaction flag is cleared. At this point, any coherence requests that were previously stalled are served. \textit{Note:} If an interrupt (e.g., for a context switch) arrives during this phase, it is postponed until the TSHR-to-cache writes are completed, which typically requires only a few clock cycles.

    \item \textbf{Abort Phase:} The transaction is aborted. All TSHRs are invalidated, an abort signal is sent to the processor, and the active transaction flag is cleared. \textit{Note:} The abort-triggering event may have occurred either before the \texttt{Store-Conditional} or during the exclusivity acquisition phase.
\end{itemize}

Transitions between these states are triggered by memory instructions such as \texttt{Load-Linked}, \texttt{Store}, and \texttt{Store-Conditional}, as well as by abort-triggering events such as coherence requests targeting the transaction’s read/write-set or processor context switches. The transaction's behavior also depends on whether it holds the token at the time of the event.

\begin{figure}[H]
    \centering
    \includegraphics[width=1\textwidth]{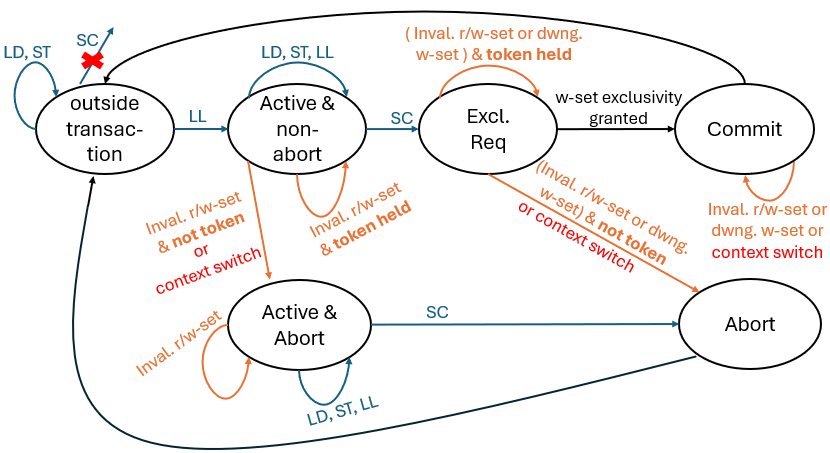}
    \caption{Token-Based Transaction Execution FSM
}

\label{fig:fsm}
\end{figure}

Figure~\ref{fig:fsm} illustrates this state machine. Blue transitions represent processor-issued memory instructions, while orange transitions represent abort-triggering events.

\textbf{Note:} For the \textit{Sorted and Sequential Exclusivity Requests} technique, the overall FSM structure remains the same, with two key modifications:
(1) Transitions that depend on whether the token is held are removed, since the mechanism does not rely on token-based priority.
(2) The \textit{Exclusivity Acquisition Phase} is broken down into $n$ smaller states, where $n$ is the number of cache lines in the transaction’s write-set.

\begin{figure}[H]
    \centering
    \includegraphics[width=1\textwidth]{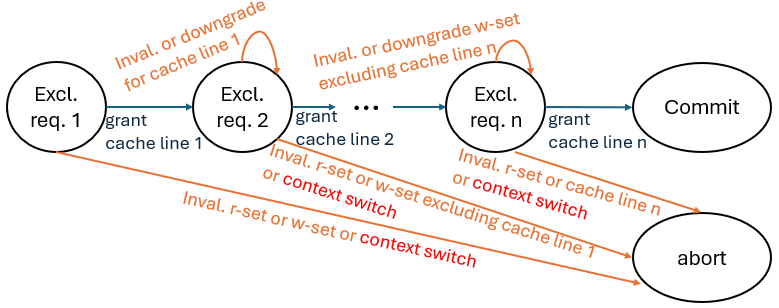}
    \caption{FSM for Ordered Write-Set Exclusivity Acquisition
}

\label{fig:fsm_sorted}
\end{figure}

Figure~\ref{fig:fsm_sorted} illustrates this refinement by decomposing the single exclusivity acquisition state into $n$ sequential states, each corresponding to a step in the ordered acquisition of one write-set cache line.

\paragraph{Recovery from Context Switch.}
When an interrupt occurs and triggers a context switch, the system must preserve the transactional state of the thread being preempted. Specifically, the \textit{Transaction Active Flag} indicates whether a transaction was active at the moment of the interrupt. Upon resuming execution, if this flag was set—implying that the thread was in the middle of a transaction—the system must restore this state by setting the \textit{Transaction Active Flag} again and immediately updating the transaction’s status to \textbf{abort}.

This approach ensures that any upcoming memory instruction such as a \texttt{Load-Linked} will not be misinterpreted as the beginning of a new transaction. Moreover, by explicitly setting the transaction’s status to abort, we avoid the complexity of handling partially completed transactional states, speculative data, or pending coherence interactions. This simplification is justified, as such events are extremely rare—given that a transaction’s execution time in our design typically lasts only a few tens to hundreds of clock cycles.
\chapter{Simulation Using Gem5}

\section{Overview of Gem5}

The \texttt{Gem5} simulator~\cite{gem5} is an open-source, community-supported simulation framework for computer architecture research. It provides a highly modular and configurable environment, enabling detailed modeling of a wide range of system components, including CPU cores, cache-coherent memory hierarchies, interconnects, and I/O devices. All these components are fully parameterized and can be adapted to simulate a variety of system architectures.

\texttt{Gem5} can operate in two primary execution modes, depending on the level of system detail required:

\begin{itemize}
    \item \textbf{Full System (FS) Mode:} Boots a complete Linux-based operating system (e.g., Ubuntu 20.04) and simulates the entire system stack, including kernel-level activity, I/O, virtual memory, and context switching. FS mode is ideal for evaluating system-level effects in realistic scenarios.
    
    \item \textbf{System Call Emulation (SE) Mode:} Instead of simulating a full OS, \texttt{gem5} emulates Linux system calls in userspace. SE mode ignores the timing of several system-level effects such as TLB misses, page faults, and actual OS behavior. 
\end{itemize}

Regarding processor models, \texttt{gem5} provides both:
\begin{itemize}
    \item \textbf{Functional (non-timing-accurate) simulation:} Used primarily for correctness and high-speed functional validation. A typical model here is \texttt{AtomicSimpleCPU}.
    \item \textbf{Timing simulation:} Models detailed cycle-accurate 
   interactions between components, useful for microarchitectural 
   performance evaluation and for simulating realistic scenarios. 
   Representative models include \texttt{O3CPU} (out-of-order) and 
   \texttt{MinorCPU} (in-order with pipelining).

\end{itemize}

\vspace{4em}
\texttt{Gem5} offers two distinct memory system models, each suitable for different levels of fidelity and flexibility:

\begin{itemize}
    \item \textbf{Classic Memory System:} A built-in cache hierarchy with a fixed MOESI snooping coherence protocol. It is simple and fast, allowing users to create custom cache hierarchies without dealing directly with coherence logic. However, it lacks protocol-level configurability and precise modeling of coherence interactions. The classic memory system is supported across all ISAs, CPU models, and memory controllers.

    \item \textbf{Ruby Memory System:} A detailed memory system that provides user-defined cache coherence protocols via the \texttt{SLICC} language. Ruby includes detailed cache memory and coherence models, as well as a detailed network model (\texttt{Garnet}). It supports various coherence implementations and it is possible to extend it to new coherence models. Ruby is mostly a drop-in replacement for the classic memory system, though it is not fully compatible with classic gem5 caches.

\end{itemize}

\texttt{Gem5} also supports multiple Instruction Set Architectures (ISAs), including \texttt{RISC-V}, \texttt{x86}, \texttt{Arm}, and \texttt{MIPS}, making it a flexible platform for simulating a broad range of modern and legacy systems.

Finally, the simulator allows for microarchitectural modifications. Computer Architects can modify internal structures—such as L1 data caches or pipeline stages—to explore novel hardware ideas or evaluate design trade-offs.

\subsection{Reflections on Modifying Gem5}

I first want to point out that \texttt{gem5} is a great tool and, without a doubt, the most full-featured computer architecture simulator available. It is an extensive project—comprising approximately 550{,}000 lines of pure \texttt{C++} code\footnote{Estimated using a custom script that I developed to count non-empty, non-comment lines in all \texttt{.cc} and \texttt{.hh} files inside the \texttt{src} directory.}—which is expected, given that it simulates an entire computer system in software.

Despite its size, a key strength of \texttt{gem5} is that in order to implement or modify a specific hardware component, one does not need to understand the entire codebase. However, it is absolutely essential to develop a deep and complete understanding of the specific code segment you are modifying. For instance, in my case, I aimed to introduce modifications in the L1 Data Cache. This required studying "only" a subset of files within the \texttt{cache} module, but every line in those files had to be thoroughly understood—there was no room for shallow reading or skipping over lines.

Before starting, I did not expect the comprehension effort to be this demanding. However, it turned out to be a truly challenging task. The code is written in modern \texttt{C++}, using advanced software engineering techniques that fully leverage the object-oriented paradigm. This makes it harder to trace dependencies across different parts of the codebase, especially for newcomers—Thanks to \texttt{gdb}, which proved to be a lifesaver.

Additionally, the \texttt{RISC-V} architecture is not as maturely supported in \texttt{gem5} compared to \texttt{x86} or \texttt{Arm}, which introduced extra difficulties during the evaluation phase—a topic I elaborate on in the next subsection.

\subsection{Simulating Multi-Threaded Programs in SE Mode}

A key challenge in \texttt{SE}-mode simulation is how to support multi-threaded applications in the absence of an actual operating system kernel—which provides essential services like thread scheduling and memory management. 

In the case of the \texttt{x86} architecture, \texttt{gem5} provides robust support for system calls and basic multi-threading. It internally emulates threading-related system calls, enabling multi-threaded applications to execute by mapping each thread to a different CPU, thereby simulating parallel execution.

Unfortunately, for \texttt{RISC-V}, support for multi-threading and thread-related system calls in \texttt{SE}-mode is incomplete. This made simulating multi-threaded programs under RISC-V another challenging task.

Inspired by the solution implemented for \texttt{x86}, I modeled each thread as a separate \texttt{gem5} process and mapped it to a different simulated CPU. However, \texttt{gem5}'s \texttt{SE}-mode simulates separate virtual address spaces (i.e., distinct page tables) per process, which meant there was no shared physical memory between them.

To solve this, I manually allocated a shared physical memory region and mapped it into the virtual address space of each process by modifying their page tables. This allowed processes to communicate via shared memory while maintaining separate page tables.

Based on this configuration, threads (i.e., simulated processes) had to allocate memory only from the shared region to ensure visibility across other processes. To manage this safely, I implemented a custom memory allocator (\texttt{malloc}) that assigned a distinct chunk of the shared region to each thread, avoiding overlapping memory usage while maintaining inter-process visibility.

\subsection{Limitations of the Classic gem5 Memory System}

In order to implement the proposed token-based mechanism, which guarantees forward progress under contention, the memory system must support delayed coherence responses. Specifically, an \texttt{L1 Data Cache} must be able to \textit{defer responding} to \texttt{Invalidate} requests for cache lines in the transaction's read/write-set and \texttt{Downgrade} requests for cache lines in the write-set.

As previously discussed, the classic memory system in \texttt{gem5} implements a built-in MOESI snooping protocol. Unfortunately, this coherence model introduces a key limitation: when a cache requests exclusive access to a cache line, the system broadcasts invalidation messages to all other caches that hold the cache line. However, acknowledgments (\texttt{Ack}) are only required from the cache (if any) that holds the line in the \texttt{Modified} state. Caches holding the cache line in the \texttt{Shared} state are not required to respond before the requesting cache is granted exclusive ownership.

This behavior imposes a critical constraint that prevents implementing the token-based mechanism within the classic memory system, as the mechanism relies on the ability of the \texttt{L1 Data Cache} serving the transaction that holds the token to delay responses to invalidation requests for its read/write-set—even when those lines are in the \texttt{Shared} state. The classic model’s lack of support for stalling such invalidations breaks the assumptions of the token-based priority scheme.

\textit{Note (for future investigation):} While writing this paragraph, I am considering that this limitation does not inherently prevent the implementation of the \textit{sorted and sequential exclusivity request} technique in the classic memory system. This is because delayed responses are only required for write-set cache lines that have already been acquired in \texttt{Exclusive} state—not for lines in \texttt{Shared} state. Therefore, it may be feasible to simulate that technique within the classic system, particularly when evaluating multi-word atomic read-modify-write operations, such as \texttt{DCAS}.

\section{Modifications to Gem5}

To simulate the behavior of our architectural design—excluding repeated attempts and without implementing any mechanisms to guarantee forward progress—modifications were applied exclusively to a minimal set of \texttt{gem5} source files. Specifically, I altered the following:

\begin{itemize}
    \item \texttt{gem5/src/mem/cache/base.hh/.cc}: This file pair implements the core base class that provides fundamental functionality for all cache types.
    \item \texttt{gem5/src/mem/cache/cache.hh/.cc}: This file pair inherits from the base class and adds higher-level functionality specific to generic cache objects.
\end{itemize}

In order to assess the scale of the modifications introduced, I developed a simple script that counts the number of non-comment, non-empty lines of code in each of the modified files. This provided an accurate quantification of the effective code changes.

The table ~\ref{table:loc_diff} presents the clean line counts for both the modified files and their original base versions, highlighting the differences introduced during development.

\renewcommand{\arraystretch}{1.2}
\begin{table}[H]
\centering
\begin{tabular}{c||c|c|c|c}
   & \textbf{Base Version} & \textbf{Modified Version} & \textbf{\# Line Difference} \\
\hline
\hline
\texttt{base.cc}  & 1753 & 2183 & +430 \\
\hline
\texttt{base.hh}  &  520 &  586 &  +66 \\
\hline
\texttt{cache.cc} &  744 &  888 & +144 \\
\hline
\texttt{cache.hh} &   57 &   58 &   +1  \\
\end{tabular}
\caption{Comparison of Clean Lines of Code  Before and After Modification}

\label{table:loc_diff}
\end{table}

It is important to note that \texttt{gem5} does not define a dedicated class for L1 Data Caches. Instead, all cache instances—data or instruction—are objects of the same \texttt{Cache} class. To restrict the newly introduced functionality exclusively to the L1 Data Cache, conditional logic was added throughout the codebase to ensure the new mechanisms only apply when the active cache instance corresponds to the L1 Data Cache.

Beyond these four files, the only additional modification was made to the Load/Store Queue (that allows outstanding reads and writes):

\begin{itemize}
\item \texttt{gem5/src/cpu/minor/lsq.cc}: I removed a condition that prevented a \texttt{Store-Conditional} instruction from issuing a memory access if the target address did not match the address of the last \texttt{Load-Linked}. While this check aligns with the classical semantics of SC instructions, it had to be relaxed to support our generalized transactional model.
\end{itemize}

In summary, I only modified two \texttt{.cc} files—\texttt{base.cc} and \texttt{cache.cc}—which are directly related to cache behavior and, by extension, to the L1 Data Cache. No changes were made to other components, such as packet structures (which carry coherence-related information) or CPU internals. This selective modification highlights that our implementation introduces hardware changes exclusively at the level of the L1 Data Cache. 

Approximately 650 lines of clean, functional code were added to the \texttt{gem5} codebase to support the proposed mechanism. I estimate that with a significantly more polished implementation—leveraging modern C++ features and following stricter code reuse and modularity principles—this number could potentially be reduced by half, to around 300 lines. This estimate does not account for the auxiliary code required to support repeated attempts in the transactional model.

\chapter{Programming Examples and Benchmarks}

While previous chapters have focused on the internal design of the mechanism, it is equally important to understand how it can be applied in practice. This chapter bridges that gap by demonstrating how the mechanism can be employed in real-world scenarios, clarifying its intended usage from a programmer’s perspective.

In the following sections, we provide a detailed explanation of the programming examples that are used to evaluate the performance of our implementation and test its correctness. We also describe the expected transactional behaviors, taking into account the potential congestion they may introduce. This analysis serves as a foundation for interpreting the results presented in the next chapter.

\section{Microbenchmarks}
This category includes benchmarks that, while not representative of complete applications, are essential for evaluating the behavior of our implementation under controlled conditions. These microbenchmarks allow us to deliberately construct scenarios—such as high contention or transaction overlap—that help reveal the limitations and performance boundaries of the proposed mechanism. As such, they are a powerful tool for exploring how the system responds to edge cases and stress situations.

\subsection{Short-Duration Counting Benchmarks}

In this category, each of the $n$ concurrent threads attempts to atomically increment all $k$ shared counters (each placed in a separate cache line) as a unit, a total of $2^{13} / n$ times, with $n$ ranging from 2 to 8 and $k$ from 2 to 4. That is, each increment operation affects all $k$ counters simultaneously and atomically---either all are incremented, or none are. This operation can be viewed as a \textit{multi-word fetch-and-add}, where the update to all $k$ counters occurs as a single transaction. These transactions are very short: each involves $k \times 2$ memory accesses (a load and a store per counter), and a total of $k \times 3$ instructions (load, add, and store per counter). Because all transactions access the exact same memory locations---namely, the shared counters---contention among threads is extremely high, representing a worst-case synchronization scenario.

Figure~\ref{fig:short-counting-thread} illustrates the high-level logic executed by a single thread, which repeatedly attempts to atomically update two shared counters until $2^{13} / n$ successful transactions have been completed. The actual implementation of the transactional update—performed using the proposed mechanism within an inline assembly block—is shown in Figure~\ref{fig:increment2-inline-asm}.

\begin{figure}[H]
\centering
\begin{minipage}{0.8\textwidth}
\rule{\linewidth}{1pt}
\begin{minted}[fontsize=\small, breaklines, baselinestretch=1.1]{c}
successes = 0;
while (successes < (8192 / n) ) {
    commit = increment_2_counters_atomically(
                 shared_counter1,
                 shared_counter2,
                 added_value1,
                 added_value2);
    if (commit) successes++;
    else backoff();
}
\end{minted}
\rule{\linewidth}{1pt}
\end{minipage}
\caption{Thread logic for performing $2^{13} / n$ successful atomic updates on two shared counters. 
If a transaction aborts, the thread performs a short random back-off (2–5 cycles) using \texttt{rand()}. 
This helps diversify execution patterns and avoids identical scenarios in simulation.}
\label{fig:short-counting-thread}
\end{figure}

\noindent
\textbf{Inline assembly for precise transactional control.}
Our proposed mechanism relies on the use of RISC-V atomic instructions \texttt{load-linked (lr)} and \texttt{store-conditional (sc)}, which are not directly accessible through standard C code. Therefore, inline assembly is used to provide precise and low-level control over their execution within the implementation. This approach allows us to bypass potential compiler optimizations or language-level abstractions that could interfere with the strict ordering and semantics of atomic memory operations. By embedding assembly directly within C code, we maintain compatibility with the simulator infrastructure while accurately modeling the hardware-level behavior of transactions.

In the implementation shown in Figure~\ref{fig:increment2-inline-asm}, placeholders such as \texttt{\%0}, \texttt{\%1}, etc.\ appear inside the assembly block. These refer to C variables passed into the assembly statement and are linked to operand constraints—such as \texttt{"r"}—which instruct the compiler to allocate those variables into general-purpose registers. This mechanism enables seamless integration between the C environment and the underlying instruction-level semantics of RISC-V.

\begin{figure}[H]
\centering
\begin{minipage}{0.95\textwidth}
\rule{\linewidth}{1pt}
\begin{minted}[fontsize=\small, breaklines, baselinestretch=1.1]{c}
int increment_2_counters_atomically(uint32_t *counter1,
                                    uint32_t *counter2,
                                    uint32_t added_value1,
                                    uint32_t added_value2)
{
    int return_value = 0;
    asm volatile (
        "    lw     t4, 0(%2)         \n"  
        "    lw     t5, 0(%3)         \n"  
        /* Begin Transaction */
        "    lr.w   t0, 0(%0)         \n"  // add counter1 to read-set
        "    lr.w   t1, 0(%1)         \n"  // add counter2 to read-set
        "    add    t0, t0, t4        \n"  
        "    sw     t0, 0(%0)         \n"  // add updated counter1 to write-set
        "    add    t0, t1, t5        \n"  
        "    sc.w   t3, t0, 0(%1)     \n"  // add updated counter2 to write-set
        /* End Transaction */
        "    sw     t3, 0(%4)         \n"  // Save SC outcome (commit or abort)
        :
        : "r"(counter1), "r"(counter2),
          "r"(&added_value1), "r"(&added_value2),
          "r"(&return_value)
    );

    return (!return_value);
}
\end{minted}
\rule{\linewidth}{1pt}
\end{minipage}
\caption{Atomic update of two shared counters using the proposed mechanism. 
Note: The \texttt{sc.w} (store-conditional) instruction is the only memory instruction with three operands. 
Its first operand (\texttt{t3}) is a register that stores the result of the operation: 0 if the store succeeded and 1 if it failed. 
In our case, this value indicates whether the transaction was successfully committed.}

\label{fig:increment2-inline-asm}
\end{figure}

\subsection{Long-Duration Counting Benchmarks}

This benchmark category is a direct extension of the \textit{Short Counting Benchmarks}, preserving the same core behavior: $n$ concurrent threads atomically increment $k$ shared counters (each placed in a separate cache line), with each thread performing $2^{13} / n$ increments. As in the short counting benchmark, $n$ ranges from 2 to 8 and $k$ ranges from 2 to 4. The key difference lies in the lifetime of each transaction.

In the long-duration version of the benchmark, transactions are deliberately extended by inserting several \texttt{noop} (no operation) instructions within their body. These \texttt{noop} instructions are inserted \textit{after} the read-set has been fully established via \texttt{load-linked} operations on all counters. Therefore, they do not alter the transaction’s read/write-set, but simply increase the time window during which a conflict with another transaction may be detected. As a result, longer-duration transactions have a higher probability of detecting a conflict before they reach their commit point.

The purpose of the \textit{Long-Duration Counting Benchmarks} is to simulate the same congestion scenarios as the short ones, while observing how the system behaves when transactions stay active for a longer period of time. In a realistic scenario, this could correspond to transactions that include additional non-memory instructions such as conditional branches (\texttt{if} statements), arithmetic logic, or control-flow structures like loops—factors that increase transaction duration without necessarily increasing memory pressure.

\section{Concurrent Data Structures}

This section presents the concurrent data structures developed for the evaluation of our implementation. In addition to describing these structures, it also aims to illustrate the underlying programming model and the design considerations involved in building parallel programs that leverage the proposed mechanism effectively.

\subsection{Producer/Consumer Queue (FIFO) Benchmark}

In this benchmark, a total of $n$ concurrent threads are used, where $n$ ranges from 2 to 16. Half of the threads ($n/2$) act as producers, while the other half act as consumers. The benchmark terminates after a total of $2^{13}$ (8192) operations, consisting of $2^{12}$ (4096) enqueues and $2^{12}$ (4096) dequeues, evenly distributed among producer and consumer threads. Thus, each producer thread performs $2^{12} / (n/2)$ (i.e., $8192 / n$) enqueue operations, and each consumer thread performs $2^{12} / (n/2)$ (i.e., $8192 / n$) dequeue operations.

The queue is implemented as a singly linked list with two pointers: \texttt{head}, which points to the beginning of the list and serves as the point where \texttt{dequeue} operations are performed, and \texttt{tail}, which points to the end of the list where new elements are inserted via \texttt{enqueue} operations.

Before describing the actual code for these operations, we first analyze four key scenarios that the implementation must handle correctly to ensure atomicity.

\paragraph{\textbf{1. Two or more concurrent enqueues while \texttt{sizeof(queue)} > 0}.}

If two concurrent transactions attempt to enqueue simultaneously, both will read the current value of the \texttt{tail} pointer in order to locate the last node of the queue. Each transaction then proceeds to update the \texttt{tail\_node->next} field to point to its own \texttt{new\_node}, and also updates the global \texttt{tail} pointer to refer to this \texttt{new\_node}, which now becomes the new tail node.

\begin{figure}[ht]
    \centering
    \includegraphics[width=1\textwidth]{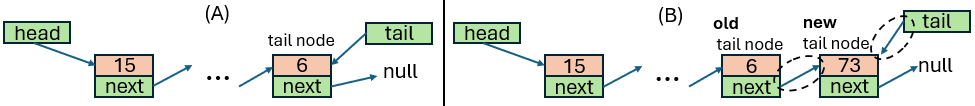}
\caption{
It is important to note that each \texttt{enqueue} transaction must atomically update both the \texttt{tail} pointer and the \texttt{tail\_node->next} field. Therefore, its write-set includes the cache lines corresponding to these two memory locations.
Additionally, at the beginning of the transaction, the \texttt{tail} pointer is part of the read-set, as it is read to determine the current tail node. However, since this pointer is also updated later during the transaction, it transitions into the write-set. As a result, by the time the transaction reaches the \texttt{sc} (store-conditional) instruction, its effective read-set may appear empty. This minimizes the risk of read-set invalidation.}

    \label{fig:2enqueues_size>1}
\end{figure}

This scenario is illustrated in Figure~\ref{fig:2enqueues_size>1}, where subfigure (A) shows the state of the queue before any \texttt{enqueue} occurs, and subfigure (B) shows the queue after a successful \texttt{enqueue(73)} operation.

\paragraph{\textbf{2. Two or more concurrent dequeues while \texttt{sizeof(queue)} > 1}.}

In this case, multiple concurrent transactions attempt to perform \texttt{dequeue} operations while the queue contains more than one element. All transactions read the \texttt{head} pointer and prepare to remove the current head node by updating the \texttt{head pointer} to point to \texttt{head\_node->next}. 

\begin{figure}[ht]
    \centering
    \includegraphics[width=1\textwidth]{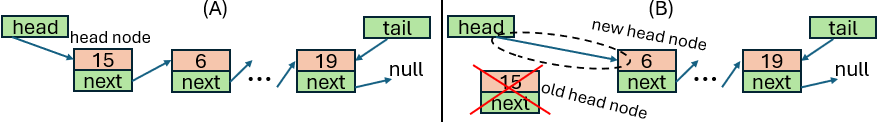}
\caption{
Each \texttt{dequeue} transaction reads the \texttt{head} pointer to locate the current head node and then reads \texttt{head\_node->next} to identify the next node, which will become the new head node. It subsequently updates the \texttt{head} pointer to point to this next node.
As a result, the transaction’s read-set includes the cache lines of both \texttt{head} and \texttt{head\_node->next}, while its write-set consists solely of the cache line containing the \texttt{head} pointer. This minimal write-set—of size one—makes the scenario favorable under contention: although multiple transactions may attempt to dequeue simultaneously, only one will acquire exclusive access to the \texttt{head} cache line, and thus only that transaction can successfully commit. Since no other memory locations are contested, at least one transaction is always guaranteed to make forward progress.
}

    \label{fig:2dequeues_size>1}
\end{figure}

This scenario is illustrated in Figure~\ref{fig:2dequeues_size>1}, where subfigure (A) shows the state of the queue before any \texttt{dequeue} occurs, and subfigure (B) shows the queue after a successful \texttt{dequeue} operation.

\paragraph{Transition to mixed-contention scenarios.}

In the previous two scenarios, we focused on the cases where \texttt{sizeof(queue) > 0} for \texttt{enqueue} operations and \texttt{sizeof(queue) > 1} for \texttt{dequeue} operations. This allowed us to isolate contention within each operation type: contention occurred only among enqueuers in the \texttt{enqueue} scenario and only among dequeuers in the \texttt{dequeue} scenario.

We now shift our attention to mixed-contention scenarios, where both enqueuers and dequeuers may simultaneously access and modify overlapping parts of the queue. Specifically, we analyze the behavior when \texttt{enqueue} is invoked on an empty queue (\texttt{sizeof(queue) = 0}) and when \texttt{dequeue} is invoked on a queue of size one. In these cases, producers and consumers are no longer isolated and may contend for the same nodes or pointers, introducing new concurrency challenges.

\paragraph{3. Concurrent \texttt{enqueue} and \texttt{dequeue} on a single-element queue.}

When the queue contains exactly one element (\texttt{sizeof(queue) = 1}), a \texttt{dequeue} operation attempts to update both the \texttt{head} and \texttt{tail} pointers to \texttt{NULL}, effectively making the queue empty. At the same time, an \texttt{enqueue} operation may attempt to update the \texttt{next} pointer of that sole node to point to the newly allocated node and the \texttt{tail} pointer to reflect the new end of the queue.

This introduces a potential race condition: both operations may access and attempt to modify overlapping parts of the queue (i.e., the tail pointer). Therefore, when designing the \texttt{enqueue} and \texttt{dequeue} algorithms, special care must be taken to ensure atomicity in such edge cases.

\begin{figure}[ht]
    \centering
    \includegraphics[width=0.9\textwidth]{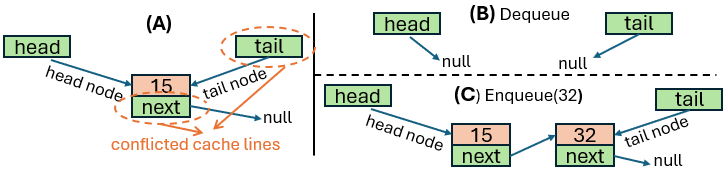}
\caption{
}
    \label{fig:enqueue_dequeue_1}
\end{figure}

Figure~\ref{fig:enqueue_dequeue_1} illustrates this situation. Subfigure~(A) shows the queue when it contains a single node. Subfigure~(B) depicts the queue after a successful \texttt{dequeue}, and Subfigure~(C) shows the result after a successful \texttt{enqueue}(32).

\noindent
\textbf{This is a critical case that highlights the importance of understanding how the proposed mechanism ensures atomicity.}  
In a queue with exactly one element—where \texttt{head\_node == tail\_node}—a concurrent \texttt{enqueue} and \texttt{dequeue} must be resolved such that only one of them commits successfully, while the other aborts. This scenario allows us to fully explore the behavior and correctness of our transactional model.

The detailed access patterns of each operation are as follows:

\begin{itemize}
  \item \textbf{The \texttt{dequeue} transaction:}
  \begin{enumerate}
    \item Reads the \texttt{head} pointer to identify the current head node.
    \item Reads the \texttt{head\_node->next} pointer to determine the new head node.
    \item Observes that \texttt{head\_node->next == NULL}, inferring that the queue will become empty.
    \item Proceeds to write to both the \texttt{head} and \texttt{tail} pointers to set them to \texttt{NULL}.
  \end{enumerate}

  \item \textbf{The \texttt{enqueue} transaction:}
  \begin{enumerate}
    \item Reads the \texttt{tail} pointer to identify the current tail node.
    \item Attempt to update both the \texttt{tail\_node->next} and the \texttt{tail} pointer to point to the newly created node.
  \end{enumerate}
\end{itemize}

As a result:
\begin{itemize}
  \item The \texttt{dequeue}'s \textbf{write-set} is \{\texttt{head pointer}, \texttt{tail pointer}\}, and its \textbf{read-set} includes \texttt{sole\_node->next}.
  \item The \texttt{enqueue}'s \textbf{write-set} is \{\texttt{tail pointer}, \texttt{sole\_node->next}\}.
\end{itemize}

This leads to a conflict on the \texttt{tail pointer} and the \texttt{next} field of the sole node. As a result, transactional conflict detection naturally enforces atomicity: at most one transaction can commit, while the other will detect the conflict and abort. No additional conditional logic or explicit synchronization is required—correct behavior is guaranteed as long as all relevant cache lines are properly included in the transactional read and write sets.

\paragraph{4. Concurrent \texttt{enqueue} and \texttt{dequeue} on an empty queue.}

When the queue is empty, an \texttt{enqueue} operation must update both the \texttt{head} and \texttt{tail} pointers to point to the newly inserted node, effectively making it the sole element in the list. Conversely, a \texttt{dequeue} operation reads the \texttt{head} pointer, detects that it is \texttt{NULL}, and therefore determines that the queue is empty—terminating without attempting to write anything.

In the case of a concurrent \texttt{enqueue} and \texttt{dequeue} on an empty queue, a subtle interaction may occur: the \texttt{enqueue}'s \textbf{write-set} overlaps with the \texttt{dequeue}'s \textbf{read-set} through the shared \texttt{head} pointer. This means that although the \texttt{dequeue} does not write any data, it can still cause the \texttt{enqueue} to abort if a specific condition occurs. In particular, if the \texttt{enqueue} transaction has already acquired exclusive access to the \texttt{head} pointer (but not yet the \texttt{tail} pointer), and a concurrent \texttt{dequeue} reads the \texttt{head} pointer, a downgrade request will be triggered. This downgrade will lead the \texttt{enqueue} transaction to detect a conflict and abort.

\textit{Note:} If the \texttt{enqueue} had already acquired both the \texttt{head} and \texttt{tail} pointers in exclusive state, the downgrade would have been delayed until the transaction completed, allowing it to commit successfully. Therefore, this is the only specific interleaving in which a \texttt{dequeue} can cause an \texttt{enqueue} to abort.

Moreover, since our proposed mechanism requires all transactions to end with a \texttt{store-conditional (sc)} instruction, even such read-only transactions must include a terminating \texttt{sc} operation. In this case, a dummy \texttt{sc} can be inserted that does not affect the state of the queue regardless of whether it succeeds or fails. One way to achieve this is by performing a \texttt{sc} on a local variable that is private to the thread.

\begin{figure}[ht]
    \centering
    \includegraphics[width=0.6\textwidth]{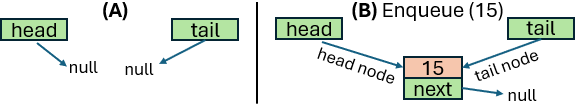}
\caption{Enqueue on an empty queue. The \texttt{head} and \texttt{tail} pointers are both updated to reference the newly inserted node.}

    \label{fig:enqueue_empty}
\end{figure}

Figure~\ref{fig:enqueue_empty} illustrates this case. Subfigure~(A) shows the queue in its empty state, while subfigure~(B) shows the result after a successful \texttt{enqueue}(15).

Now that we have thoroughly discussed the key scenarios that must be considered when designing the \texttt{enqueue} and \texttt{dequeue} operations, we proceed to present the actual implementation code for these functions. As with previous examples, the implementation is written using \textit{inline assembly} to directly employ the RISC-V \texttt{Load-Linked/Store-Conditional} instructions, which form the foundation of our proposed mechanism.

The structure of each queue node used in the implementation is shown in Figure~\ref{fig:node-struct}.

\begin{figure}[H]
\centering
\begin{minipage}{0.8\textwidth}
\rule{\linewidth}{1pt}
\begin{minted}[fontsize=\small, breaklines, baselinestretch=1.1]{c}
typedef struct node {
    uint32_t* data;             // Pointer to data
    uint32_t* padding[7];       // Dummy space
    struct node* next;          // Pointer to next node
} node_t;
\end{minted}
\rule{\linewidth}{1pt}
\end{minipage}
\caption{Structure definition of a queue node. The dummy padding is used to ensure that the \texttt{next} pointer resides on a different cache line than the \texttt{data} field. This separation is used to simulate more general scenarios in which the \texttt{data} and \texttt{next} fields may belong to different cache lines.}
\label{fig:node-struct}
\end{figure}

\noindent
Figures~\ref{fig:enqueue-inline-asm} and~\ref{fig:dequeue-inline-asm} show the implementations of the \texttt{enqueue} and \texttt{dequeue} operations, respectively, using the proposed mechanism. These implementations were integrated into our simulation framework.

\begin{figure}[H]
\centering
\begin{minipage}{0.8\textwidth}
\rule{\linewidth}{1pt}
\begin{minted}[fontsize=\small, breaklines, baselinestretch=1.1]{c}
int enqueue(node_t* new_node){
    int return_value=1;
    node_t **head_ptr = HEAD_PTR_ADDR;
    node_t **tail_ptr = TAIL_PTR_ADDR;
    asm volatile(
        "lw t0, 0(%0) \n"       // t0 = new_node
        /* Begin Transaction */
        "lr.w t2, 0(%1) \n"     // add tail_ptr to read-set
        "beqz t2, L1f \n"        // if *tail_ptr == NULL go to L1
        "sw t0, 64(t2) \n"      // tail_node->next = new_node
        "j L2f \n"               // go to L2
        "L1: \n"
        "sw t0, 0(%3) \n"       // *head_ptr = new_node
        "L2: \n"
        "sc.w t4, t0, 0(%1) \n" // *tail_ptr = new_node
        /* End Transaction */
        "sw t4, 0(%2) \n"       // save sc outcome (commit or abort)
        : 
        : "r"(&new_node), "r"(tail_ptr),
          "r"(&return_value), "r"(head_ptr)
    );
    return !return_value;
}
\end{minted}
\rule{\linewidth}{1pt}
\end{minipage}
\caption{The transaction begins with a \texttt{load-linked (lr.w)} on the \texttt{tail pointer}, which adds it to the read-set. If the \texttt{tail pointer} is \texttt{NULL}, this indicates that the queue is empty, and the transaction must also update the \texttt{head pointer} to point to the new node—corresponding to the \texttt{enqueue on empty queue} scenario. If the queue is not empty, only the \texttt{tail pointer} and the \texttt{next} field of the current tail node are written. The \texttt{store-conditional (sc.w)} instruction stores 0 (success) or 1 (failure) in register \texttt{t4}, which is then saved to a local variable \texttt{return\_value} to indicate whether the transaction committed successfully.}

\label{fig:enqueue-inline-asm}
\end{figure}

\begin{figure}[H]
\centering
\begin{minipage}{0.95\textwidth}
\rule{\linewidth}{1pt}
\begin{minted}[fontsize=\small, breaklines, baselinestretch=1.1]{c}
int dequeue(){
    int ret_value = 1;
    int dequeued_value = -1;
    node_t **head_ptr = HEAD_PTR_ADDR;
    node_t **tail_ptr = TAIL_PTR_ADDR;
    asm volatile(
        /* Begin Transaction */
        "lr.w t2, 0(%0) \n"         // add head_ptr to read-set
        "beqz t2, L1f \n"            // if (*head_ptr == NULL), go to L1
        "lw t4, 0(t2)   \n"         // non-transactional load of head_node
        "lw t3, 0(t4)   \n"         // non-transactional load of head_node->data
        "sw t3, 0(%3)   \n"         // store the dequeued value
        "addi t5, t2, 64 \n"     // t5 = address of the 'next' field of the head node
        "lr.w t0, 0(t5) \n"       // add head_node->next to read-set
        "beqz t0, L2f  \n"         // if (head_node->next == NULL) go to L2
        "sc.w t1, t0, 0(%0) \n"   // set *head_ptr = head_node->next
        /* End Transaction Point 1, case: >1 nodes */
        "j L3f \n"                   // go to L3

        "L1: \n"                     // queue is empty
        "li t0, 1 \n"               // load value 1 into t0
        "sc.w t1, t0, 0(%2) \n"     // dummy sc to end transaction
        /* End Transaction Point 2, case: empty queue */
        "j L3f \n"

        "L2: \n"                     // queue has one element
        "li t0, 0 \n"                // load value 0 (NULL) into t0
        "sw t0, 0(%1) \n"           // set *tail_ptr = NULL
        "sc.w t1, t0, 0(%0) \n"     // set *head_ptr = NULL
        /* End Transaction Point 3, case: single element queue */

        "L3: \n"
        "sw t1, 0(%2) \n"     // in any case save sc outcome (abort or commit)
        : 
        : "r"(head_ptr), "r"(tail_ptr), "r"(&ret_value), "r"(&dequeued_value)
    );
         // Success: dequeued_value != -1
         // Empty queue: dequeued_value == -1 && ret_value == 0
         // Otherwise: transaction aborted
    return dequeued_value;
}
\end{minted}
\rule{\linewidth}{1pt}
\end{minipage}
\caption{Dequeue Implementation Using the Proposed Mechanism.}
\label{fig:dequeue-inline-asm}
\end{figure}

\noindent{}
The implementation of the \texttt{dequeue} operation begins with a \texttt{load-linked (lr.w)} instruction to load the \texttt{head} pointer, thereby adding it to the transaction’s \textit{read-set}. The value of this pointer is then checked: if it is \texttt{NULL}, the queue is empty. In this case, a dummy \texttt{store-conditional (sc.w)} instruction is issued on a local variable to formally terminate the transaction. At this point, the \texttt{dequeued\_value} remains \texttt{-1}, as initially assigned. 

\textit{Note:} This dummy \texttt{sc} may fail, so the correctness of the dequeue logic does not depend on whether it succeeds or not.

If the queue is not empty, the transaction proceeds with two non-transactional \texttt{loads (lw)} instructions: one to read the \texttt{head\_node} and one to read its \texttt{data} field. These are intentionally left outside the transactional domain, since the values they read are immutable within the context of a \texttt{dequeue}—no other thread will write to them. This design choice saves valuable TSHR slots.

The \texttt{data} field is then written into a local variable (\texttt{dequeued\_value}), adding this store to the \textit{write-set}. Next, the address of the \texttt{head\_node->next} field is computed based on the known layout of the node structure and read using a second \texttt{load-linked (lr.w)}, thereby adding it to the \textit{read-set}.

If \texttt{head\_node->next == NULL}, it means the queue contains only a single element. In that case, both the \texttt{head} and \texttt{tail} pointers must be updated to \texttt{NULL}, and this is done using a \texttt{sw} on the \texttt{tail} and a \texttt{sc.w} on the \texttt{head}.  Otherwise, the queue has more than one node, and the \texttt{head} pointer is simply updated to point to the next node.

In all cases, only one of the possible \texttt{sc.w} instructions is executed, and its result is stored in a shared local variable to indicate whether the transaction committed successfully.

At the end, we can interpret the transaction outcome based on the state of the \texttt{dequeued\_value} and the \texttt{return\_value}:
\begin{itemize}
    \item If \texttt{dequeued\_value} is different from \texttt{-1}, the transaction has committed and returned a valid value.
    \item If \texttt{dequeued\_value == -1}, but the transaction committed, this implies an empty queue.
    \item If neither occurred, the transaction aborted.
\end{itemize}

In our implementation, we treat dequeueing from an empty queue as a failed operation, although it could be extended to return richer status information if needed.

\subsection{Sorted Doubly Linked-List Benchmark}

In this benchmark, we have $n$ threads, where $n$ ranges from 2 to 16, each performing $2^{12}/n$ insertions, followed by $2^{12}/n$ deletions of the same nodes once the insertions are completed.
To ensure contention among threads---preventing each thread from operating exclusively in disjoint node neighborhoods of the list---we appropriately adjusted the data values of the nodes inserted by each thread. Specifically, thread~0 inserts nodes with data values:
$1000 + 0 \cdot n$, $1000 + 1 \cdot n$, $\ldots$, $1000 + ((2^{12}/n) - 1) \cdot n$;
thread~1 inserts nodes with data values:
$1001 + 0 \cdot n$, $1001 + 1 \cdot n$, $\ldots$; and so on.

In general, thread~$t$ inserts nodes with data values of the form:
\[
1000 + t + i \cdot n \quad \text{for} \quad i = 0, 1, \ldots, \left(\frac{2^{12}}{n} - 1\right)
\]
It is important to highlight that operations on sorted doubly linked lists represent a class of concurrent programs distinct from those covered by the previously discussed programs. More specifically, the neighborhoods of nodes where an insert or delete will take place are not known in advance. Therefore, each such operation requires a traversal of the list to locate the appropriate neighborhood. In our implementation, this search is performed using an optimistic search. Once the target neighborhood is located, we retain pointers to the relevant nodes, and then begin a transaction to apply the necessary modifications. At the beginning of the transaction, we perform a validation step to verify whether the state of the neighborhood remains unchanged. If it does not, the transaction completes without performing any modifications, and the operation is retried.

\begin{figure}[ht]
    \centering
    \includegraphics[width=0.8\textwidth]{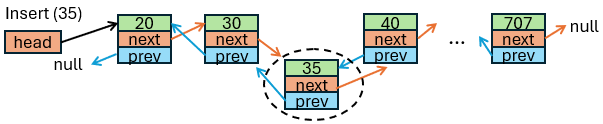}
    \caption{State after an insert operation. The new node is linked between two existing nodes, and the corresponding pointers are updated.}
    \label{fig:insert}
\end{figure}

\begin{figure}[H]
    \centering
    \includegraphics[width=0.75\textwidth]{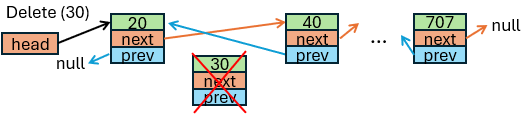}
    \caption{State after a delete operation. The node is removed, and adjacent nodes are re-linked to maintain list integrity.}
    \label{fig:delete}
\end{figure}

Before proceeding to the implementation, we illustrate in Figures~\ref{fig:list},~\ref{fig:insert}, and~\ref{fig:delete} the structure and state transitions of the list. Figure~\ref{fig:list} shows a typical doubly linked list; Figure~\ref{fig:insert} shows the state after an insertion; and Figure~\ref{fig:delete} shows the state after a deletion.

\begin{figure}[ht]
    \centering
    \includegraphics[width=0.7\textwidth]{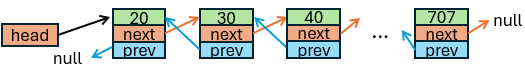}
    \caption{Structure of a doubly linked list. Each node contains pointers to both the previous and the next node.}
    \label{fig:list}
\end{figure}

\noindent
We now analyze the special cases that must be carefully handled in the implementation of \texttt{insert} and \texttt{delete} operations in a sorted doubly linked list in order to ensure atomicity.

\vspace{1em}

\begin{figure}[ht]
    \centering
    \includegraphics[width=0.7\textwidth]{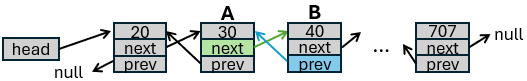}
    \caption{Read/write-sets for concurrent insertions between the same neighboring nodes $A$ and $B$. Both transactions access and attempt to modify the same pointers, leading to a conflict that allows only one to commit successfully.}
    \label{fig:insert_betw}
\end{figure}

\noindent
\textbf{Concurrent insertions between the same neighboring nodes (A and B).}
In this scenario, two (or more) threads attempt to insert nodes between the same pair of neighboring nodes $A$ and $B$. As shown in Figure~\ref{fig:insert_betw}, both threads try to modify the \texttt{next} pointer of $A$ and the \texttt{prev} pointer of $B$. Therefore, it suffices that these two cache lines are included in the read/write-set of each transaction, ensuring that at most one thread will successfully commit while the others will abort.

\begin{figure}[ht]
    \centering
    \includegraphics[width=0.5\textwidth]{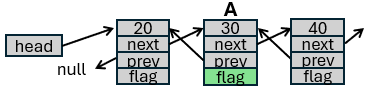}
   \caption{Conflict on the \texttt{flag} field during concurrent deletions of the same node $A$.}

    \label{fig:delete_node}
\end{figure}

\noindent
\textbf{Concurrent deletions of the same node.}
In this scenario, two (or more) threads attempt to delete the same node A. In our implementation, the thread that deletes a node must set the node's \texttt{flag} field to 1. Therefore, if all threads include the \texttt{flag} field in their read/write-set while it is still 0, at most one thread will succeed in changing its value to 1. 
Figure~\ref{fig:delete_node} illustrates the \texttt{flag} field of the node as a distinct cache line.

\begin{figure}[ht]
    \centering
    \includegraphics[width=0.6\textwidth]{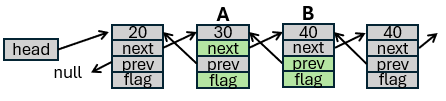}
    \vspace{0.5em}
    \textit{Insert A-B}
    
    \vspace{0.3em}
    
    \includegraphics[width=0.6\textwidth]{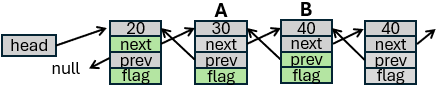}
    \vspace{0.3em}
    \textit{Delete A}
    
    \vspace{0.3em}
    
    \includegraphics[width=0.6\textwidth]{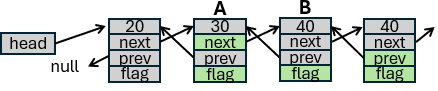}
    \vspace{0.5em}
    \textit{Delete B}
    
   \caption{Read/write-sets in concurrent insertion and deletion within a node neighborhood, leading to transactional conflicts due to overlapping accesses.}

    \label{fig:insert_delete_conflict}
\end{figure}

\noindent
\textbf{Concurrent insertion between nodes A and B and deletion of node A or B.}
In this scenario, one thread attempts to insert a node between nodes $A$ and $B$, while another thread concurrently attempts to delete either node $A$ or node $B$. Figure~\ref{fig:insert_delete_conflict} illustrates the cache lines involved in the read/write-sets of the corresponding transactions. The top image shows the read/write-set of the insertion between nodes $A$ and $B$. The middle image shows the read/write-set for the deletion of node $A$, and the bottom image shows the read/write-set for deletion of node $B$.

As shown, at most one of these transactions can successfully commit. This is because the insertion transaction includes the \texttt{flag} fields of both nodes $A$ and $B$ in its read-set. Therefore, if either of these nodes is deleted before the insertion commits, the insertion transaction will detect a conflict and abort. Conversely, if the insertion commits first, it will cause the deletion transaction to abort. Specifically, the insertion writes to the \texttt{prev} pointer of node $B$, which will invalidate the transaction that attempts to delete node $A$, and it writes to the \texttt{next} pointer of node $A$, which will invalidate the transaction that attempts to delete node $B$.

\begin{figure}[H]
    \centering    
    \includegraphics[width=0.6\textwidth]{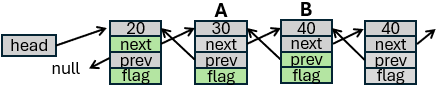}
    \vspace{0.3em}
    \textit{Delete A}
    
    \vspace{0.3em}
    
    \includegraphics[width=0.6\textwidth]{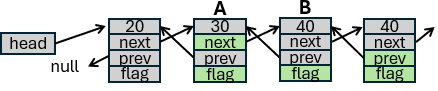}
    \vspace{0.5em}
    \textit{Delete B}
    
    \caption{Read/write sets in the concurrent deletion of adjacent nodes $A$ and $B$. The overlapping access to the \texttt{flag} fields introduces transactional conflicts, allowing at most one transaction to commit.}
    \label{fig:delete_conflict}
\end{figure}

\noindent
\textbf{Concurrent deletion of adjacent nodes A and B.}
In this scenario, two (or more) threads concurrently attempt to delete two adjacent nodes, $A$ and $B$, such that $A \rightarrow \texttt{next} = B$ and $B \rightarrow \texttt{prev} = A$.
Figure~\ref{fig:delete_conflict} illustrates the read/write sets involved in these transactions. The top image shows the transaction that deletes node $A$, and the bottom image shows the transaction that deletes node $B$.

At most one of these transactions can successfully commit. The transaction that deletes node $A$ writes the \texttt{flag} field of node $A$ to 1, which causes the transaction attempting to delete node $B$ to abort, since it includes the \texttt{flag} of node $A$ in its read-set. Conversely, the transaction that deletes node $B$ writes to the \texttt{flag} of node $B$, invalidating the transaction that deletes node $A$, which includes the \texttt{flag} of node $B$ in its read-set. This mutual interference leads to a transactional conflict, ensuring that only one of the deletions can complete successfully, thereby preserving consistency in the list.

\noindent
The core ideas when designing these operations can be summarized as follows:
\begin{itemize}
    \item \textbf{Perform an optimistic search} to locate the neighborhood of nodes that must be modified, whether for an insertion or a deletion.

    \item \textbf{Perform validation at the beginning of each transaction} to ensure that the state of the identified node neighborhood has not changed between the search phase and the start of the transaction.

    \item \textbf{Include the appropriate set of cache lines in the transaction's read/write set} so that non-independent operations will conflict and cannot commit concurrently.

    \item \textbf{Minimize the size of the read/write set} to allow independent operations to proceed concurrently and commit successfully without unnecessary conflicts.
\end{itemize}

\noindent
We now present the implementations of insertion and deletion operations on a sorted doubly-linked list using the proposed mechanism.

The structure of the doubly-linked list node used in our implementation is shown in Figure~\ref{fig:list-struct}.

\begin{figure}[H]
\centering
\begin{minipage}{0.8\textwidth}
\rule{\linewidth}{1pt}
\begin{minted}[fontsize=\small, breaklines, baselinestretch=1.1]{c}
typedef struct node {
    uint32_t* data;             // Pointer to data
    uint32_t* padding[7];       // Dummy space
    struct node* next;          // Pointer to next node
    uint32_t* padding[7];       // Dummy space
    struct node* prev;          // Pointer to prev node
    uint32_t* padding[7];       // Dummy space
    uint32_t* flag;             // Pointer to flag
} node_t;
\end{minted}
\rule{\linewidth}{1pt}
\end{minipage}
\caption{Structure definition of a doubly-linked list node. Padding is used to ensure that each field is placed on a separate cache line, simulating scenarios where fields may reside in different cache lines.}
\label{fig:list-struct}
\end{figure}

\paragraph{Insertion in a Sorted Doubly Linked List.}
We now present the implementation for inserting a node into a sorted doubly linked list. The process is structured into two parts: an initial optimistic search phase, followed by one of four insertion cases depending on the position where the new node must be placed.

In the optimistic search phase, we traverse the list to locate the appropriate insertion point. Specifically, we identify two neighboring nodes $A$ and $B$ such that the new node should be inserted between them. These two pointers, stored as \texttt{prev\_node} and \texttt{next\_node}, can either be \texttt{NULL} or valid node addresses, giving rise to the following four cases:

\begin{description}
  \item[Case 1:] $A = \texttt{NULL}$ and $B = \texttt{NULL}$: the list is empty. We begin a transaction and validate that the \texttt{head\_pointer} is still \texttt{NULL}. If successful, we update it to point to the new node; otherwise, we abort via a dummy \texttt{store-conditional}.

    \item[Case 2:] $A = \texttt{NULL}$ and $B \neq \texttt{NULL}$: insertion at the head. We validate that the current head still satisfies \texttt{head\_node->data > new\_node->data} and that the head node is not logically deleted (\texttt{flag == 0}). If so, we update both the \texttt{prev} pointer of the current head and the \texttt{head\_pointer} itself, and set the \texttt{next} field of the new node to point to the old head node.

  \item[Case 3:] $A \neq \texttt{NULL}$ and $B = \texttt{NULL}$: insertion at the tail. We validate that $A$ is still the last node (i.e., $A->next == \texttt{NULL}$) and has not been deleted ($A->flag == 0$). If validation passes, we set $A->next = \texttt{new\_node}$ and $new\_node->prev = A$.

  \item[Case 4:] $A \neq \texttt{NULL}$ and $B \neq \texttt{NULL}$: insertion in the middle. We validate that $A->next == B$ and that neither $A$ nor $B$ has been deleted. If validation is successful, we update the pointers $A->next$, $B->prev$, and the respective fields of the new node to insert it between $A$ and $B$.
\end{description}

In all cases, the transaction's read-set includes the \texttt{flag} fields of any neighboring nodes involved. This ensures that concurrent deletions of those nodes will lead to transaction aborts, preserving correctness.

\begin{center}
\begin{minipage}{0.95\textwidth}
\rule{\linewidth}{1pt}
\begin{minted}[fontsize=\small, breaklines, baselinestretch=1.1]{c}
// Common preprocessing: determine where to insert
int dll_insert_sorted(node_t* new_node){
    int ret_value = 1;
    node_t **head_ptr = HEAD_PTR_ADDR;
    int new_data = *(new_node->data);
    int new_flag = *(new_node->flag);
    node_t *next_node = *head_ptr;
    node_t *prev_node = NULL;

    while (next_node != NULL && *(next_node->data) < new_data) {
        prev_node = next_node;
        next_node = next_node->next;
    }
\end{minted}
\rule{\linewidth}{1pt}
\label{fig:insert-preprocess}
\end{minipage}
\end{center}

\vspace{1em}

\begin{center}
\begin{minipage}{0.95\textwidth}
\rule{\linewidth}{1pt}
\begin{minted}[fontsize=\small, breaklines, baselinestretch=1.1]{c}
// Case 1: Empty list (both prev_node and next_node are NULL)
if (prev_node == NULL && next_node == NULL) {
    asm volatile(
        "lw t0, 0(%0) \n"             // t0 = new_node

        /* Begin Transaction */
        "lr.w t2, 0(%1) \n"           // Load *head_ptr into t2 (add to read-set)
        
        /* Validation step: Begin */
        "bnez t2, 1f \n"              // If head is not NULL, jump to abort path
        /* Validation step: End */
        
        "sc.w t4, t0, 0(%1) \n"       // Attempt to set *head_ptr = new_node
        /* End Transaction */

        "sw t4, 0(%2) \n"             // Store result (0: success, 1: failure)
        "j 2f \n"

        /* Abort Path */
        "1: \n"
        "li t4, 1 \n"                      // set t4 = 1
        "sc.w t5, t4, 0(%2) \n"       // Dummy SC to complete transaction
        "sw t4, 0(%2) \n"             // Set ret_value = 1 (Failure)

        "2: \n"
        :
        : "r"(&new_node), "r"(head_ptr), "r"(&ret_value)
    );
}

\end{minted}
\rule{\linewidth}{1pt}
\label{fig:insert-empty}
\end{minipage}
\end{center}

\vspace{1em}

\begin{center}
\begin{minipage}{0.95\textwidth}
\rule{\linewidth}{1pt}
\begin{minted}[fontsize=\small, breaklines, baselinestretch=1.1]{c}
// Case 2: Insert at head (prev_node == NULL, next_node != NULL)
else if (prev_node == NULL && next_node != NULL) {
    asm volatile(
        "lw t0, 0(%0) \n"             // t0 = new_node
        "lw t1, 0(%3) \n"             // t1 = new_data

        /* Begin Transaction */
        "lr.w t2, 0(%1) \n"           // t2 = *head_ptr (old head)
        "lw t6, 0(t2) \n"             // t6 = old_head->data pointer
        "lw t5, 192(t2) \n"           // t5 = &old_head->flag
        "lr.w t4, 0(t5) \n"           // t4 = old_head->flag
        
        /* Validation step: Begin */
        "lw t5, 0(t6) \n"             // t5 = old_head->data
        "blt t5, t1, 1f \n"           // If old_head->data < new_data, abort
        "bnez t4, 1f \n"              // If old_head->flag != 0, abort
        /* Validation step: End */

        "sw t0, 128(t2) \n"           // old_head->prev = new_node
        "sw t2, 64(t0) \n"            // new_node->next = old_head
        "sc.w t4, t0, 0(%1) \n"       // *head_ptr = new_node
        /* End Transaction */

        "sw t4, 0(%2) \n"             // Store SC result to ret_value
        "j 2f \n"

        /* Abort Path */
        "1: \n"
        "li t4, 1 \n"                      // Set t4 = 1 
        "sc.w t5, t4, 0(%2) \n"       // Dummy SC to close transaction
        "sw t4, 0(%2) \n"             // ret_value = 1 (Failure)

        "2: \n"
        :
        : "r"(&new_node), "r"(head_ptr), "r"(&ret_value), "r"(&new_data)
    );
}

\end{minted}
\rule{\linewidth}{1pt}
\label{fig:insert-head}
\end{minipage}
\end{center}

\vspace{1em}

\begin{center}
\begin{minipage}{0.95\textwidth}
\rule{\linewidth}{1pt}
\begin{minted}[fontsize=\small, breaklines, baselinestretch=1.1]{c}
// Case 3: Insert at tail (prev_node != NULL, next_node == NULL)
else if (prev_node != NULL && next_node == NULL) {
    asm volatile(
        "lw t0, 0(%0) \n"             // t0 = new_node

        /* Begin Transaction */
        "lw t2, 0(%1) \n"             // t2 = prev_node
        "addi t1, t2, 64 \n"          // t1 = &prev_node->next
        "lw t4, 192(t2) \n"           // t4 = &prev_node->flag
        "lr.w t3, 0(t1) \n"           // t3 = prev_node->next
        "lr.w t5, 0(t4) \n"           // t5 = prev_node->flag

        /* Validation step: Begin */
        "bnez t3, 1f \n"              // If prev_node->next != NULL, abort
        "bnez t5, 1f \n"              // If prev_node->flag != 0, abort
        /* Validation step: End */
        

        "sw t2, 128(t0) \n"           // new_node->prev = prev_node
        "sc.w t4, t0, 0(t1) \n"       // prev_node->next = new_node
        /* End Transaction */

        "sw t4, 0(%2) \n"             // Store SC result to ret_value
        "j 2f \n"

        /* Abort Path */
        "1: \n"
        "li t4, 1 \n"                      // Set t4 = 1
        "sc.w t5, t4, 0(%2) \n"       // Dummy SC to complete transaction
        "sw t4, 0(%2) \n"             // ret_value = 1 (Failure)

        "2: \n"
        :
        : "r"(&new_node), "r"(&prev_node), "r"(&ret_value), "r"(&new_data)
    );
}

\end{minted}
\rule{\linewidth}{1pt}
\label{fig:insert-tail}
\end{minipage}
\end{center}

\vspace{1em}

\begin{center}
\begin{minipage}{0.95\textwidth}
\rule{\linewidth}{1pt}
\begin{minted}[fontsize=\small, breaklines, baselinestretch=1.1]{c}
// Case 4: Insert in the middle (both prev_node and next_node != NULL)
else {
    asm volatile(
        "lw t0, 0(%0) \n"             // t0 = new_node

        /* Begin Transaction */
        "lw t2, 0(%1) \n"             // t2 = prev_node
        "lw t3, 0(%4) \n"             // t3 = next_node

        "addi t1, t2, 64 \n"          // t1 = &prev_node->next
        "lr.w t4, 0(t1) \n"           // t4 = prev_node->next
        /* Validation step: Begin */
        "bne t3, t4, 1f \n"           // If prev_node->next != next_node, abort

        "lw t6, 192(t2) \n"           // t6 = &prev_node->flag
        "lr.w t6, 0(t6) \n"           // t6 = prev_node->flag
        "bnez t6, 1f \n"              // If prev_node->flag != 0, abort

        "lw t6, 192(t3) \n"           // t6 = &next_node->flag
        "lr.w t6, 0(t6) \n"           // t6 = next_node->flag
        "bnez t6, 1f \n"              // If next_node->flag != 0, abort
        /* Validation step: End */
        "sw t3, 64(t0) \n"            // new_node->next = next_node
        "sw t0, 128(t3) \n"           // next_node->prev = new_node
        "sw t2, 128(t0) \n"           // new_node->prev = prev_node
        "sc.w t5, t0, 0(t1) \n"       // prev_node->next = new_node
        /* End Transaction */
        "sw t5, 0(%2) \n"             // Store SC result to ret_value
        "j 2f \n"
        /* Abort Path */
        "1: \n"
        "li t4, 1 \n"                       // Set t4 = 1
        "sc.w t5, t4, 0(%2) \n"       // Dummy SC to complete transaction
        "sw t4, 0(%2) \n"             // ret_value = 1 (Failure)
        "2: \n"
        :
        : "r"(&new_node), "r"(&prev_node), "r"(&ret_value),
          "r"(&new_data), "r"(&next_node)
    );
}

\end{minted}
\rule{\linewidth}{1pt}
\label{fig:insert-middle}
\end{minipage}
\end{center}

\vspace{3em}

\paragraph{Deletion in a Sorted Doubly Linked List.}
We now present the implementation for deleting a node from a sorted doubly linked list. The process is structured into two parts: an initial optimistic search phase, followed by one of three deletion cases depending on the node’s location within the list.

During the optimistic search phase, the list is traversed to locate the node containing the specified data. Throughout this traversal, we maintain two pointers: \texttt{prev\_node} and \texttt{next\_node}, where \texttt{next\_node} is the candidate for deletion.

\begin{description}
  \item[Case 1: Empty list or node not found.] 
  If the list is empty or the desired node is not found, the function returns failure immediately. This case requires no transaction, as the operation is trivially invalid.

  \item[Case 2: Node to delete is the head node.]
  In this case, the node to delete is the first in the list. We begin a transaction and validate that the \texttt{head\_pointer} still points to the same node and that the node’s \texttt{flag} field is zero. If this validation fails, the transaction is completed with a dummy \texttt{store-conditional} and returns failure. Otherwise, we proceed by:
  \begin{itemize}
    \item setting the node’s \texttt{flag} to 1 to mark it deleted,
    \item updating \texttt{head\_pointer} to point to \texttt{head\_node->next},
    \item if \texttt{head\_node->next != NULL}, we set its \texttt{prev} field to \texttt{NULL},
    \item and we include the \texttt{flag} field of \texttt{head\_node->next} in the transaction’s read-set, to ensure the transaction aborts if a concurrent delete affects the next node.
  \end{itemize}

  \item[Case 3: Node is a middle or tail node.]
  In this case, the node to be deleted is either in the middle or the end of the list. We begin a transaction and validate:
  \begin{itemize}
    \item the current node’s \texttt{flag} is 0,
    \item the \texttt{prev\_node != NULL}, and the \texttt{prev\_node->flag == 0},
    \item if \texttt{next\_node->next != NULL}, then \texttt{next\_node->next->flag == 0}.
  \end{itemize}
  If validation succeeds, we:
  \begin{itemize}
    \item mark the current node’s \texttt{flag} as 1,
    \item update \texttt{prev\_node->next} to skip the node being deleted and point to \texttt{next\_node},
    \item update \texttt{next\_node->prev} to skip the node and point to \texttt{prev\_node}.
  \end{itemize}
  The transaction includes the flag fields of the adjacent nodes in the read-set. This ensures that if any of them is concurrently deleted, the transaction will abort.
\end{description}

\begin{center}
\begin{minipage}{0.95\textwidth}
\rule{\linewidth}{1pt}
\begin{minted}[fontsize=\small, breaklines, baselinestretch=1.1]{c}
// Common preprocessing: locate the node to delete
int delete_from_dllist(uint32_t data){
    int ret_value = 1;
    node_t **head_ptr = HEAD_PTR_ADDR;
    node_t *next_node = *head_ptr;
    node_t *prev_node = NULL;

    while (next_node != NULL && *(next_node->data) < data) {
        prev_node = next_node;
        next_node = next_node->next;
    }
    
    if (next_node == NULL || *(next_node->data) != data) {
        // Case 1: List is empty or data not found
        return 0;
    }

\end{minted}
\rule{\linewidth}{1pt}
\label{fig:delete-search}
\end{minipage}
\end{center}

\begin{center}
\begin{minipage}{0.95\textwidth}
\rule{\linewidth}{1pt}
\begin{minted}[fontsize=\small, breaklines, baselinestretch=1.1]{c}
// Case 2: Deleting the first node (head)
else if (next_node != NULL && prev_node == NULL) {
    asm volatile(
        "lw t0, 0(%2) \n"             // t0 = data for deletion
        "lr.w t2, 0(%0) \n"           // t2 = *head_ptr (old head)
        "lw t1, 0(t2) \n"             // t1 = old_head_node->data pointer
        "lw t5, 0(t1) \n"             // t5 = *old_head_node->data
        /* Validation step: Begin */
        "bne t5, t0, 1f \n"           // Abort if data mismatch
        
        "lw t5, 192(t2) \n"          // t5= &current->flag
        "lr.w t5, 0(t5) \n"          //Check current node->flag
        "bnez t5, 1f \n"             // if current_node->flag != 0, abort
        
        "addi t6, t2, 64 \n"           // &old_head_node->next
        "lr.w t3, 0(t6) \n"           // t3 = old_head_node->next
        "beqz t3, 3f \n"              // If t3 == NULL, skip update

        "lw t5, 192(t3) \n"           // t5 = &next->flag
        "lr.w t5, 0(t5) \n"           // Check next->flag
        "bnez t5, 1f \n"              // if next_node->flag != 0, abort
        /* Validation step: End */
        "li t4, 0 \n"                     // set t4 = 0
        "sw t4, 128(t3) \n"           // next->prev = NULL
        "3: \n"
        "li t4, 1 \n"
        "lw t5, 192(t2) \n"            // get &old_head->flag
        "sw t3, 0(%0) \n"             // head_ptr = old_head->next
        "sc.w t3, t4, 0(t5) \n"       // mark old_head as deleted
        "sw t3, 0(%1) \n"             // store SC result to ret_value
        "j 2f \n"

        "1: \n"
        "li t4, 1 \n"                      // set t4 = 1
        "sc.w t5, t0, 0(%2) \n"       // Dummy SC to complete transaction
        "sw t4, 0(%1) \n"             // failure
        "2: \n"
        :
        : "r"(head_ptr), "r"(&ret_value), "r"(&data)
    );
}
\end{minted}
\rule{\linewidth}{1pt}
\label{fig:delete-head}
\end{minipage}
\end{center}

\begin{center}
\begin{minipage}{0.95\textwidth}
\rule{\linewidth}{1pt}
\begin{minted}[fontsize=\small, breaklines, baselinestretch=1.1]{c}
// Case 3: Deleting a middle or tail node
else if (next_node != NULL && prev_node != NULL) {
    asm volatile(
        "lw t0, 0(%2) \n"
        "lw t1, 0(%0) \n"             // node to delete
        "addi t2, t1, 64 \n"          // &node->next
        "addi t3, t1, 128 \n"         // &node->prev
        "lw t6, 0(t1) \n"             // node->data
        "lw t4, 0(t6) \n"             // *node->data
        "bne t0, t4, 1f \n"           // data mismatch
        "lw t5, 192(t1) \n"           // node->flag
        "lr.w t4, 0(t5) \n"           // *node->flag
        "bnez t4, 1f \n"             // if node->flag!=0, abort
        "lr.w t4, 0(t2) \n"           // node->next
        "lr.w t5, 0(t3) \n"           // node->prev
        "beqz t5, 1f \n"              // if node->prev==NULL, abort
        "lw t6, 192(t5) \n"           // node->prev->flag
        "lr.w t2, 0(t6) \n"           // t2 = *node->prev->flag
        "bnez t2, 1f \n"              // if node->prev->flag!=0, abort
        "beqz t4, 3f \n"            // we are deleting a tail node
        "lw t3, 192(t4) \n"         // node->next->flag
        "lr.w t2, 0(t3) \n"         // *node->next->flag
        "bnez t2, 1f \n"            // if node->next->flag!=0, abort
        "sw t5, 128(t4) \n"           // node->next->prev = node->prev
        "3: \n"
        "sw t4, 64(t5) \n"            // node->prev->next = node->next
        "4: \n"
        "lw t5, 192(t1) \n"           // node->flag
        "li t4, 1 \n"                    // set t4 = 1
        "sc.w t3, t4, 0(t5) \n"       // mark node deleted
        "sw t3, 0(%1) \n"           // store SC result to ret_value
        "j 2f \n"
        "1: \n"
        "li t4, 1 \n"                      // set t4 = 1
        "sc.w t5, t0, 0(%2) \n"       // Dummy SC to complete transaction
        "sw t4, 0(%1) \n"            // ret_value = 1 (Failure)
        "2: \n"
        :
        : "r"(&next_node), "r"(&ret_value), "r"(&data),
          "r"(head_ptr), "r"(&prev_node)
    );
}
\end{minted}
\rule{\linewidth}{1pt}

\label{fig:delete-middle-tail}
\end{minipage}
\end{center}

\chapter{Simulation Results and Evaluation}

In the previous chapter, we fully described the microbenchmarks and programming examples developed for our simulation experiments. In this chapter, we present the results obtained from these simulations.

\section{Simulation Model}

For the simulation experiments, we used the system configuration summarized in Table~\ref{tab:system-model}, running in \textit{syscall emulation (SE)} mode in Gem5. Each processor is a single-issue, in-order core (RiscvMinorCPU) and is equipped with a private 64\,KB L1 data cache, which is 8-way associative and augmented with 8 Transaction Status Holding Registers (TSHRs). Though single-issue and in-order, the processor model includes an aggressive, single-cycle non-memory IPC.

Additionally, we introduced explicit timing for the commit phase: each write buffered in the TSHRs incurs a latency of \textbf{one cycle} when written to the L1 data cache.

\begin{table}[H]
\centering
\caption{System model parameters}
\label{tab:system-model}
\begin{tabular}{|l||l|}
\hline
\textbf{} & \textbf{System Model Settings} \\
\hline \hline
Processors & 2\,GHz, single-issue, in-order (RiscvMinorCPU), non-memory IPC = 1 \\
\hline
L1 Data Cache & 64\,KB, 8-way associative, 1 cycle latency, 8 TSHRs \\
\hline
L2 Cache & Shared, 256\,KB, 8-way associative, 12 cycle latency \\
\hline
Cache Coherence & MOESI snooping protocol (Classic Gem5 memory system) \\
\hline
\end{tabular}
\end{table}

\noindent
\paragraph{Statistics Collection.} All statistics presented in this chapter were collected using Gem5’s internal \texttt{stats} infrastructure. This mechanism allows precise tracking of custom and architectural events within the simulator, without introducing additional timing overheads or altering the behavior of the benchmark code itself.

\section{Counting Benchmarks Results}

We begin our evaluation with the \textit{Counting Benchmarks}, which were fully described in Chapter~5. These benchmarks stress the system under high contention, with multiple threads (ranging from 2 to 8) attempting to atomically update a group of shared counters (2-4) using the proposed mechanism. This microbenchmark allows us to explore the scalability limits of our implementation and identify the level of contention at which a forward progress mechanism becomes essential.

We present results for both short- and long-duration transactions across configurations with 2 to 4 shared counters, allowing us to study the impact of read/write-set size and transaction duration on the abort rate.

The presentation of our results is structured as follows. For each of the two benchmark categories—\textit{short-duration} and \textit{long-duration} counting transactions—we first present a breakdown of abort causes and compare the results across the subcategories with 2, 3, and 4 shared counters. This comparison enables us to evaluate how increasing the read/write-set size affects the overall abort rate.

Once both benchmark categories have been individually analyzed, we proceed to compare corresponding subcategories (e.g., 2-counter short-duration vs. 2-counter long-duration) in order to isolate and assess the impact of transaction duration on the abort rate.

Finally, we compare the execution time (measured in clock cycles) for the short-duration counting transaction benchmarks across configurations with 2 to 4 shared counters, against equivalent implementations that use a traditional Test-and-Test-and-Set (TTS) lock. Specifically, we evaluate four variants: our proposed transaction-based mechanism with and without exponential backoff, and the TTS-based implementation, also with and without exponential backoff. In both cases, the backoff strategy uses identical minimum and maximum delay bounds to ensure a fair comparison. The use of exponential delay after unsuccessful attempts is motivated by prior work \cite{anderson, scott}, which has shown that TTS locks with exponential backoff substantially outperform the versions without backoff on small-scale multiprocessors. This evaluation allows us to examine, within the implementation of our own mechanism, how introducing exponential backoff after an aborted transaction affects overall performance.

\subsection{Short-Duration Counting Transactions}

\begin{table}[H]
\centering
\caption*{Benchmark Configuration Summary}
\vspace{-0.5em}
\begin{tabular}{|l|l|}
\hline
\textbf{Parameter} & \textbf{Value} \\
\hline
Number of Threads ($n$) & 2 to 8 \\
\hline
Number of Shared Counters ($k$) & 2 to 4 \\
\hline
Total Increments per Counter & $2^{13}$ \\
\hline
Successful Transactions per Thread & $2^{13}/n$, each atomically incrementing all $k$ counters \\
\hline
Transaction Body & $k$ loads + $k$ adds + $k$ stores \\
\hline
Maximum Number of TSHR Entries Used & equal to $k$ (4) \\
\hline
\end{tabular}
\end{table}

\vspace{1em}

\begin{figure}[H]
    \centering
    \begin{minipage}[b]{0.32\textwidth}
        \centering
        \includegraphics[width=\textwidth]{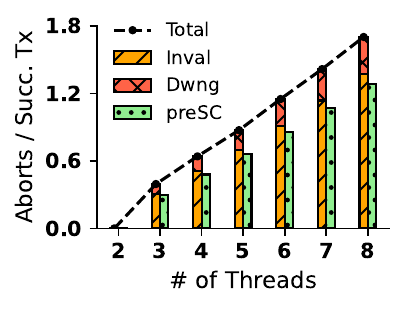}
        \caption*{2 Counters}
        \label{fig:2faa}
    \end{minipage}
    \hfill
    \begin{minipage}[b]{0.32\textwidth}
        \centering
        \includegraphics[width=\textwidth]{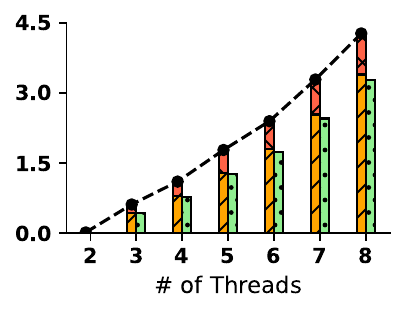}
        \caption*{3 Counters}
        \label{fig:3faa}
    \end{minipage}
    \hfill
    \begin{minipage}[b]{0.32\textwidth}
        \centering
        \includegraphics[width=\textwidth]{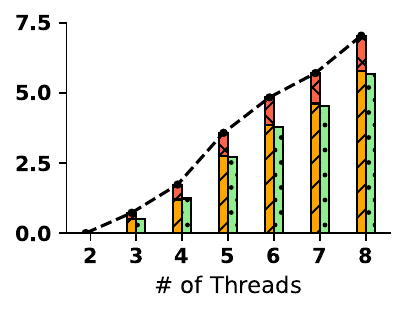}
        \caption*{4 Counters}
        \label{fig:4faa}
    \end{minipage}

\caption{
Abort ratio per successful transaction for the 2-, 3-, and 4-counter short-duration benchmarks.  
Each subplot breaks down the causes of transactional aborts:  
\textbf{Total} represents the overall number of aborts per successful transaction;  
\textbf{Inval} indicates aborts caused by cache line invalidations;  
\textbf{Dwng} indicates aborts due to cache line downgrades;  
and \textbf{preSC} captures aborts that occurred before the final store-conditional instruction was executed. \textit{Note:} If a transaction receives an \texttt{invalidate} followed later by a \texttt{downgrade}, it is categorized under failures due to \texttt{invalidate}  (i.e., we count the first abort-triggering event that occurred). 
}
\label{fig:all_faa}

\end{figure}

 \textbf{A notable observation across all benchmarks} is that the majority of aborts occur before execution reaches the store-conditional instruction. Consequently, transactions often terminate without ever issuing exclusivity requests for their write-sets, which reinforces our design choice to defer such requests until the final phase of execution. Additionally, the number of transactions that abort before reaching the store-conditional is consistently a subset of those that abort due to \texttt{invalidate}. This is expected, as no exclusivity has been acquired at that point and all accessed cache lines remain in the shared state.
 
\begin{figure}[H]
    \centering
    \includegraphics[width=0.5\textwidth]{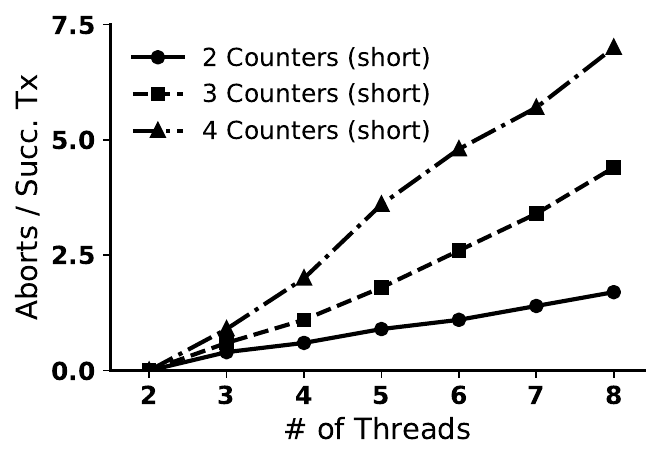}
\caption{
Abort ratio per successful transaction for the 2-, 3-, and 4-counter short-duration benchmarks, across varying thread counts. As shown, for up to 3 threads, the size of the read/write-set has a relatively minor impact on the abort ratio in these short and fast transactions. However, starting from 4 threads, increasing the read/write-set size by one leads to an almost twofold increase in the abort ratio. This trend highlights that, under high-contention scenarios (i.e., with more than 4 threads), the size of the read/write-set plays a decisive role in determining the abort ratio per successful transaction.
}
    \label{fig:comparison}
\end{figure}

\vspace{1em}
\noindent
Having presented how the abort ratio per successful transaction varies across benchmarks with different read/write-set sizes, we now bring all three cases together in a single line plot (Figure~\ref{fig:comparison}). This Combined view allows us to directly compare the impact of read/write-set size on abort behavior, particularly under high-contention scenarios.

\vspace{1em}
\subsection{Long-Duration Counting Transactions}

\begin{table}[H]
\centering
\caption*{Benchmark Configuration Summary}
\vspace{-0.5em}
\begin{tabular}{|l|l|}
\hline
\textbf{Parameter} & \textbf{Value} \\
\hline
Number of Threads ($n$) & 2 to 8 \\
\hline
Number of Shared Counters ($k$) & 2 to 4 \\
\hline
Total Increments per Counter & $2^{13}$ \\
\hline
Successful Transactions per Thread & $2^{13}/n$, each atomically incrementing all $k$ counters \\
\hline
Transaction Body & $k$ loads + $k$ adds + $k$ stores + $k \times 10$ no-ops \\
\hline
Maximum Number of TSHR Entries Used & equal to $k$ (4) \\
\hline
\end{tabular}
\end{table}

\vspace{1em}

\begin{figure}[H]
    \centering
    \begin{minipage}[b]{0.32\textwidth}
        \centering
        \includegraphics[width=\textwidth]{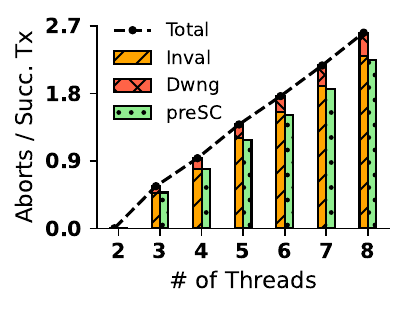}
        \caption*{2 Counters}
        \label{fig:2faa}
    \end{minipage}
    \hfill
    \begin{minipage}[b]{0.32\textwidth}
        \centering
        \includegraphics[width=\textwidth]{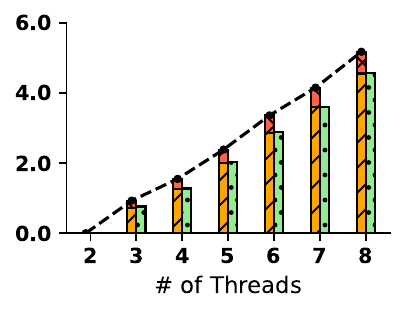}
        \caption*{3 Counters}
        \label{fig:3faa}
    \end{minipage}
    \hfill
    \begin{minipage}[b]{0.32\textwidth}
        \centering
        \includegraphics[width=\textwidth]{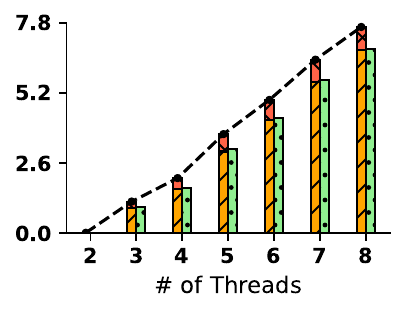}
        \caption*{4 Counters}
        \label{fig:4faa}
    \end{minipage}

 \caption{
Abort ratio per successful transaction for the 2-, 3-, and 4-counter long-duration benchmarks.  
Each subplot breaks down the causes of transactional aborts:  
\textbf{Total} represents the overall number of aborts per successful transaction;  
\textbf{Inval} indicates aborts caused by cache line invalidations;  
\textbf{Dwng} indicates aborts due to cache line downgrades;  
and \textbf{preSC} captures aborts that occurred before the final store-conditional instruction was executed. \textit{Note:} If a transaction receives an \texttt{invalidate} followed later by a \texttt{downgrade}, it is categorized under failures due to \texttt{invalidate}  (i.e., we count the first abort-triggering event that occurred). 
}
\label{fig:all_faa_slower}

\end{figure}

As is the case with the short-duration counting transaction benchmarks, the majority of aborts in the long-duration counting transactions also occur before execution reaches the store-conditional instruction. This indicates that, despite the increased transaction lifetime, most conflicts are still detected early, before any write-set exclusivity is requested.

\begin{figure}[H]
    \centering
    \includegraphics[width=0.5\textwidth]{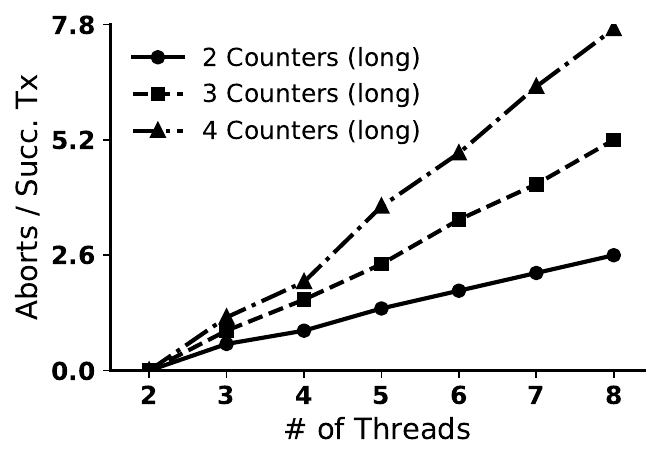}
  \caption{
Abort ratio per successful transaction for the 2-, 3-, and 4-counter long-duration benchmarks, across varying thread counts. Similar to the short-duration benchmarks, the long-duration configurations exhibit a sharp increase in abort ratio under high-contention scenarios (i.e., with more than 4 threads). Notably, increasing the read/write-set size by one results in an approximate doubling of the abort ratio per successful transaction.
}

    \label{fig:comparison_slower}
\end{figure}

\vspace{1em}
\noindent
Having presented how the abort ratio per successful transaction varies across configurations with different read/write-set sizes, we now bring all three long-duration cases together in a single line plot (Figure~\ref{fig:comparison_slower}).

\noindent
\textbf{Duration impact across configurations.} Having examined the behavior of both short- and long-duration counting benchmarks independently, we now proceed to compare corresponding subcategories across the two. Specifically, we align configurations with the same number of shared counters (e.g., 2-counter short-duration vs. 2-counter long-duration) in order to assess the impact of transaction duration on abort behavior. This comparison allows us to isolate the effect of increased transaction lifetime while keeping the read/write-set size constant, thus revealing how duration alone influences the abort ratio under varying levels of contention.

\begin{figure}[H]
    \centering
    \begin{minipage}[b]{0.32\textwidth}
        \centering
        \includegraphics[width=\textwidth]{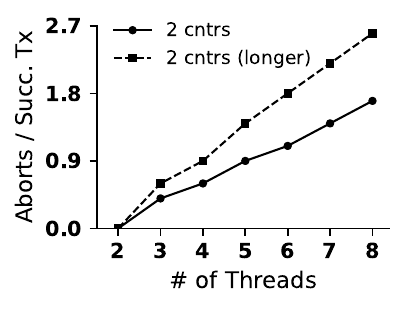}
        \caption*{2 Counters}
        \label{fig:2_short_long}
    \end{minipage}
    \hfill
    \begin{minipage}[b]{0.32\textwidth}
        \centering
        \includegraphics[width=\textwidth]{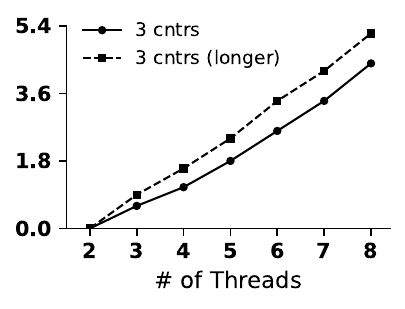}
        \caption*{3 Counters}
        \label{fig:3_short_long}
    \end{minipage}
    \hfill
    \begin{minipage}[b]{0.32\textwidth}
        \centering
        \includegraphics[width=\textwidth]{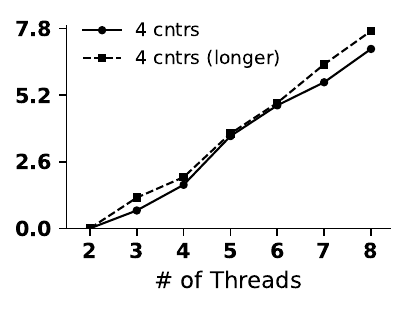}
        \caption*{4 Counters}
        \label{fig:4_short_long}
    \end{minipage}

 \caption{
Abort ratio per successful transaction for short- and long-duration benchmarks across 2-, 3-, and 4-counter configurations. Each subplot isolates the effect of transaction duration by holding the read/write-set size fixed.
}
\label{fig:short_long_all_faa}

\end{figure}

\noindent
In general, longer transaction duration leads to a higher abort ratio. However, this trend appears to be more pronounced when the read/write-set is small (i.e., in the 2-counter case). As the size of the read/write-set increases (3 or 4 counters), the influence of transaction duration on the abort ratio becomes less significant.

An important observation is that in transactions with very small read/write-sets, the duration of the transaction has a strong influence on the abort rate. In contrast, as the read/write-set grows, its size becomes the main determinant of abort behavior, while the effect of transaction duration becomes less pronounced.

\noindent
\paragraph{Comparison with Test-and-Test-and-Set (TTS).}  
In benchmarks like the one considered here—where the critical sections of all threads operate on exactly the same memory locations—a traditional locking mechanism such as Test-and-Test-and-Set (TTS) may initially seem well-suited. In these cases, contention is localized to a small set of memory addresses, rather than being distributed across the system, making simple spinlock-based synchronization relatively effective.

To evaluate how our transactional mechanism performs in such contention-heavy scenarios, we compare it against TTS using the short-duration counting transaction benchmarks. We present three execution time plots—one for each configuration with 2, 3, and 4 shared counters. Each plot reports the total execution time (in clock cycles) required to perform $2^{12}$ (4096) additions on the shared counter group. This setup allows for a direct comparison between the transactional and TTS-based implementations under identical conditions of high contention on fixed memory locations.

\begin{figure}[H]
    \centering
    \begin{minipage}[b]{0.32\textwidth}
        \centering
        \includegraphics[width=\textwidth]{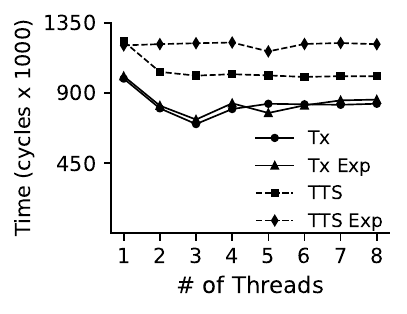}
        \caption*{2 Counters}
        \label{fig:2faacycles}
    \end{minipage}
    \hfill
    \begin{minipage}[b]{0.32\textwidth}
        \centering
        \includegraphics[width=\textwidth]{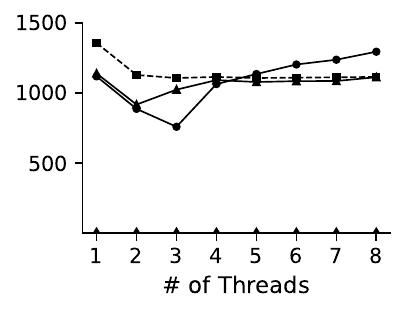}
        \caption*{3 Counters}
        \label{fig:3faacycles}
    \end{minipage}
    \hfill
    \begin{minipage}[b]{0.32\textwidth}
        \centering
        \includegraphics[width=\textwidth]{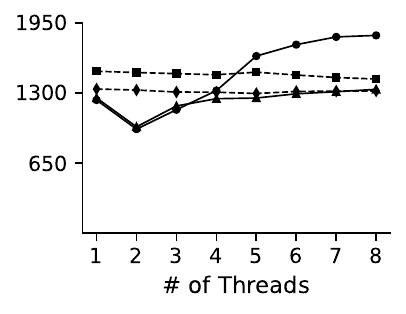}
        \caption*{4 Counters}
        \label{fig:4faacycles}
    \end{minipage}

\caption{
Execution time comparison between our transactional mechanism and a Test-and-Test-and-Set (TTS) lock across short-duration counting benchmarks with 2, 3, and 4 shared counters. Each line shows the total number of clock cycles required to complete $4096 \times k$ atomic additions, evenly distributed among $n$ threads. In each transaction, all $k$ shared counters are incremented atomically.
}

\label{fig:cycles_comparison}

\end{figure}

The results indicate that when the read/write-set size is very small—as in the 2-counter case—the implementation using the proposed mechanism outperforms the TTS-based alternative. Moreover, we observe that incorporating exponential backoff after aborted transactions does not improve performance in this scenario. As the read/write-set size increases (e.g., in the 3- and 4-counter cases), the performance of the proposed mechanism without backoff degrades significantly with higher thread counts, eventually falling behind even the TTS implementation without backoff. However, when exponential backoff is enabled, the proposed mechanism performs comparably to the TTS implementation with backoff, even under high contention—a notably positive result. In all cases, for low thread counts (i.e., fewer than 4), implementations based on the proposed mechanism consistently outperform their TTS-based counterparts.

\section{Producer/Consumer Queue (FIFO)}
In a First-In-First-Out (FIFO) queue, all threads operate either on the head or the tail node, which leads to significant contention on specific memory addresses and limits the potential for parallelism. At best, only one thread can perform an enqueue and one thread a dequeue at any given time. Because all threads access fixed memory locations, this benchmark is comparable to the shared counter benchmark in terms of contention.

However, it allows us to evaluate our mechanism in a more realistic scenario with high contention on specific memory locations. We present results from executions with varying numbers of active threads $n$, half of which act as enqueuers and the other half as dequeuers. In each execution, a total of $2^{13}$ operations are performed, evenly divided among the active threads—$2^{12}$ enqueues and $2^{12}$ dequeues.

\begin{table}[H]
\centering
\caption*{Benchmark Configuration Summary}
\vspace{-0.5em}
\begin{tabular}{|l|l|}
\hline
\textbf{Parameter} & \textbf{Value} \\
\hline
Number of Threads ($n$) & 2 to 16 (step 2): 2, 4, ..., 16 \\
\hline
Number of Enqueuers & $n/2$ \\
\hline
Number of Dequeuers & $n/2$ \\
\hline
Total Number of Operations & $2^{13}$ \\
\hline
Successful Transactions per Thread & $2^{13}/n$ \\
\hline
Maximum Number of TSHR Entries Used & 4 \\
\hline
\end{tabular}
\end{table}

\vspace{1em}

\begin{figure}[H]
    \centering
    \begin{minipage}[b]{0.48\textwidth}
        \centering
        \includegraphics[width=\textwidth]{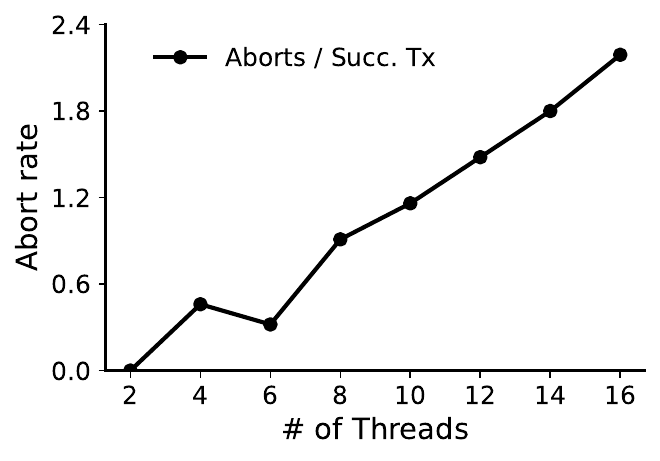}
        \caption*{}
        \label{fig:failed}
    \end{minipage}
    \hfill
    \begin{minipage}[b]{0.48\textwidth}
        \centering
        \includegraphics[width=\textwidth]{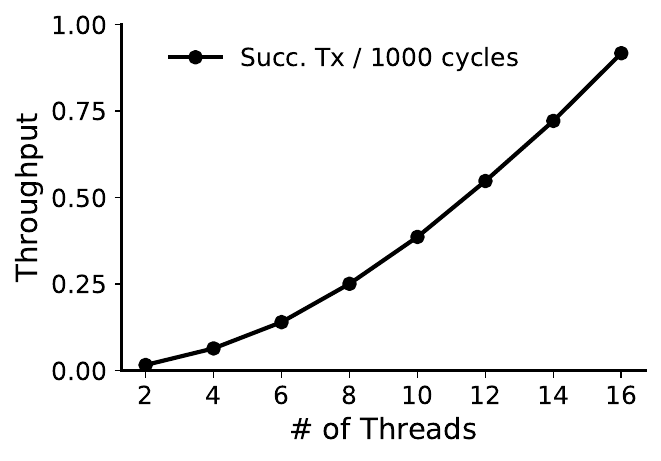}
        \caption*{}
        \label{fig:throughput}
    \end{minipage}

\caption{Two performance metrics are shown: abort rate (failed/successful transactions) on the left, and throughput (successful transactions per 1000 clock cycles) on the right.}

\label{fig:failures_throughput}

\end{figure}

\noindent
In Figure~\ref{fig:failures_throughput} we report two main metrics. The first is the abort rate, which shows how the number of aborted transactions per successful transaction increases as the number of threads grows. As expected, increasing the thread count leads to more contention (at both the head and tail of the queue), and thus a higher abort rate.

To better understand the performance implications, we also measure the throughput of successful transactions, to observe how the system's ability to complete operations evolves despite the increasing abort rate as more threads are introduced.

\noindent
From our results, we observe that despite the nearly linear increase in abort rate, increasing the number of threads leads to a superlinear speedup in the throughput of successful transactions (i.e., throughput($2X$) > throughput($X$)). This behavior typically occurs at moderate thread counts, as is the case here. However, if we were to continue increasing the number of threads—for example, to 32—it is likely that this trend would plateau, eventually hitting a scalability limit or saturation point.

\section{Sorted Doubly Linked-List}

In sorted doubly-linked lists, insertions and deletions can occur at any node in the list. This is a key difference from data structures like queues or stacks, which allows for greater parallelism. That is, multiple threads can simultaneously perform successful insertions and deletions on the list. In contrast, a queue typically allows at most one enqueuer and one dequeuer to operate at the same time. However, conflicts between threads may still arise when they modify the same region of the list, though such conflicts are expected to be less frequent.

Additionally, this benchmark introduces a new aspect not present in the previous ones: the need to locate the node to delete or to find the appropriate neighborhood for inserting a new node. This requirement involves traversing the list. Therefore, this programming example aims to show that, even though our proposed mechanism supports only limited read/write sets, it is still sufficient to implement such applications—despite their seemingly large read sets.

We evaluate our implementation by running experiments with varying numbers of active threads $n$. Each thread performs $2^{12}/n$ insertions, followed by $2^{12}/n$ deletions of the same nodes it inserted. To ensure contention among threads—preventing each from operating in a completely disjoint region of the list—we assign overlapping data ranges to the inserted nodes.

Specifically, thread $t$ inserts nodes with data values of the form $1000 + t + i \cdot n$, for $i = 0$ to $(2^{12}/n - 1)$. This offsetting scheme creates controlled overlap between threads and increases the likelihood of transactional interference during insertions and deletions.

\begin{table}[H]
\centering
\caption*{Benchmark Configuration Summary}
\vspace{-0.5em}
\begin{tabular}{|l|l|}
\hline
\textbf{Parameter} & \textbf{Value} \\
\hline
Number of Threads ($n$) & 2 to 16 \\
\hline
Number of Insertions & $2^{12}$ \\
\hline
Number of Deletions & $2^{12}$ \\
\hline
Total Number of Operations & $2^{13}$ \\
\hline
Successful Transactions per Thread & $2^{13}/n$ \\
\hline
Maximum Number of TSHR Entries Used & 8 \\
\hline
\end{tabular}
\end{table}

\noindent
Figure~\ref{fig:ddlist} presents two key metrics that characterize transactional behavior in the sorted doubly linked list benchmark:

\begin{itemize}
    \item \textbf{Abort rate:} This measures the number of aborted transactions per successful transaction.
    
   \item \textbf{Unsuccessful transactions per successful transaction:} This metric is defined as the ratio of all failed transactions—including both aborted ones and those that commit but return failure at the application level—to the number of successful transactions. Even when a transaction does not abort, it may still fail due to a validation mismatch and complete via a dummy \texttt{store-conditional}.

    Therefore, it is more accurate in such cases to count the number of \emph{unsuccessful transactions} (i.e., transactions that either aborted or completed without aborting but returned failure at the application level due to failed validation), and express their ratio relative to successful ones.

\end{itemize}

\begin{figure}[H]
    \centering
    \begin{minipage}[b]{0.7\textwidth}
        \centering
        \includegraphics[width=\textwidth]{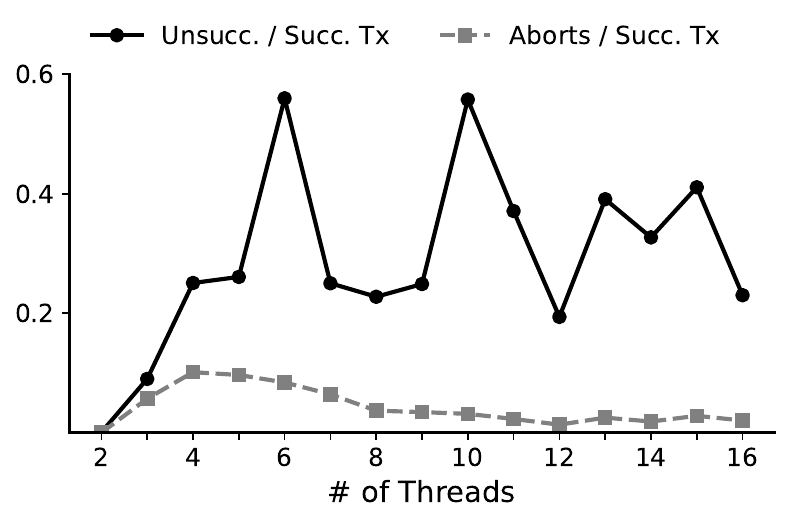}
        \caption*{}
        \label{fig:ddlist}
    \end{minipage}

\caption{Abort rate and ratio of unsuccessful to successful transactions in the sorted doubly linked list benchmark.}

\label{fig:ddlist}

\end{figure}

From the results, we observe that the majority of unsuccessful transactions were due to validation failures, while only a very small fraction—almost negligible—were due to aborts. Furthermore, the ratio of unsuccessful to successful transactions remains low across all runs, generally between 0.2 and 0.4, except for two runs where it approached 0.6. This behavior does not appear to be a function of the number of active threads, which suggests that the system remains scalable even in environments with a larger number of active threads.

As expected, our proposed mechanism performs well in scenarios where updates are not concentrated on specific nodes of the data structure, enabling high degrees of parallelism. For example, consider three nodes A, B, and C such that A→next = B and B→next = C. Our mechanism allows one thread to insert a node between A and B, while another thread simultaneously inserts a node between B and C. In contrast, in fine-grained locking implementations—where each node is protected by a separate lock—this would not be possible due to contention on node B.

\chapter{Conclusions and Future Work}

This bachelor thesis presented the design of a limited read/write-set Hardware Transactional Memory (HTM) system that does not require modifications to standard cache coherence protocols or the Instruction Set Architecture (ISA). It also introduced hardware-supported extensions to guarantee forward progress under high contention scenarios. To demonstrate the programmability of the proposed HTM mechanism, a set of custom microbenchmarks was developed, including atomic increments on multiple counters, a Producer/Consumer pattern using a First-In-First-Out (FIFO) data structure (Concurrent Queue), and a Producer/Consumer pattern on a concurrent sorted doubly-linked list. Finally, the proposed HTM was implemented in a system call emulation environment using the gem5 simulator, and its performance was evaluated using the custom microbenchmarks.

\section*{Future Work}
Several promising directions remain open for future exploration, which we were unable to address within the limited timeframe of this thesis.

\begin{enumerate}
    \item \textbf{Implementation of Forward Progress Mechanisms in gem5.} While the conceptual extensions for guaranteeing forward progress were proposed, a full implementation and evaluation of these mechanisms within the gem5 simulator would provide deeper insights into their practicality and performance impact.

    \item \textbf{Hardware Design and Validation.} Translating the proposed HTM design into an actual hardware implementation—using tools such as hardware description languages (HDLs) and FPGA-based prototyping—would allow for a more accurate assessment of its area, power, and timing characteristics.
    
    \item \textbf{Algorithmic Exploration in the New Programming Model.} Further research could focus on designing and evaluating popular parallel and distributed algorithms using the proposed transactional model. This would help better understand its expressiveness, limitations, and potential benefits across a wider range of applications.
\end{enumerate}

\section*{A Final Note}
Just before completing this thesis, I had the chance to read the 2023 retrospective on the Bulk paper by Luis Ceze, James M. Tuck, Calin Cascaval, and Josep Torrellas~\cite{ceze2023bulk}, published in the \textit{Collection of Retrospectives on Selected Papers from the Second 25 Years of the International Symposium on Computer Architecture (ISCA)}. In the final section of their retrospective (\textit{``THE FUTURE''}), the authors reflect on their original expectation—dating back to 2006—that Hardware Transactional Memory (HTM) would become a popular technique in commercial computer systems. However, looking back on this vision, they identify several reasons why such techniques did not gain widespread adoption. Notably, they remark that ``a major reason has to be the imbalance between the relatively high hardware complexity of TM implementations and the small set of existing applications that can use TM to substantially improve performance or programmability.''

This observation strongly resonated with the motivation behind this thesis. In response to that imbalance, we proposed a deliberately constrained HTM design with very low hardware complexity, targeting a meaningful subset of applications where HTM can still offer benefits in both performance and programmability.

\nocite{*}
\bibliographystyle{plain}
\bibliography{references}

\end{document}